\documentclass[sigconf]{acmart}

\AtBeginDocument{%
  \providecommand\BibTeX{{%
    \normalfont B\kern-0.5em{\scshape i\kern-0.25em b}\kern-0.8em\TeX}}}

\setcopyright{acmlicensed}
\copyrightyear{2018}
\acmYear{2018}
\acmDOI{XXXXXXX.XXXXXXX}

\acmConference[Conference acronym 'XX]{Make sure to enter the correct
  conference title from your rights confirmation emai}{June 03--05,
  2018}{Woodstock, NY}
\acmISBN{978-1-4503-XXXX-X/18/06}

\usepackage{algpseudocode}
\usepackage[ruled,vlined,linesnumbered]{algorithm2e}

\usepackage{amsmath, amsfonts}
\usepackage{textcomp}
\usepackage{graphicx}
\usepackage{balance}  
\usepackage{float}
\usepackage{supertabular}
\usepackage{caption}
\usepackage{url}
\usepackage{xcolor}
\usepackage{enumitem}
\usepackage{xspace}
\usepackage{amsthm}
\usepackage{multirow}
\usepackage{subfig}

\theoremstyle{definition}
\newtheorem{definition}{Definition}

\theoremstyle{problem}
\newtheorem*{problem}{Problem Statement}

\theoremstyle{example}
\newtheorem{example}{Example}

\theoremstyle{property}
\newtheorem{property}{Property}

\theoremstyle{lemma}

\theoremstyle{corollary}
\newtheorem{corollary}{Corollary}

\theoremstyle{observation}

\begin{document}

\title{Effective Individual Fairest Community Search over Heterogeneous Information Networks}



\author{Taige Zhao}
\affiliation{%
  \institution{Deakin University}
  \city{Geelong}
  \country{Australia}
  \orcid{0000-0001-8623-1878}
}
\email{zhaochr@deakin.edu.au}

\author{Jianxin Li} 
\authornote{Both authors contributed equally to this research.}

\affiliation{%
  \institution{Deakin University}
  \city{Geelong}
  \country{Australia}
}
\email{jianxin.li@deakin.edu.au}

\author{Ningning Cui}
\affiliation{%
  \institution{Chang'an University}
  \city{Xi'an}
  \country{China}
}
\email{csnncui@chd.edu.cn}

\author{Wei Luo}
\affiliation{%
  \institution{Deakin University}
  \city{Geelong}
  \country{Australia}
}
\email{wei.luo@deakin.edu.au}

\begin{abstract}
Community search over heterogeneous information networks has been applied to wide domains, such as activity organization and team formation. 
From these scenarios, the members of a group with the same treatment often have different levels of activity and workloads, which causes unfairness in the treatment between active members and inactive members (called individual unfairness).
%
However, existing works do not pay attention to individual fairness and do not sufficiently consider the rich semantics of HINs (e.g., high-order structure), which disables complex queries.
To fill the gap, we formally define the issue of individual fairest community search over HINs (denoted as IFCS), which aims to find a set of vertices from the HIN that own the same type, close relationships, and small difference of activity level and has been demonstrated to be NP-hard.
To do this, we first develop an exploration-based filter that reduces the search space of the community effectively.
Further, to avoid repeating computation and prune unfair communities in advance, we propose a message-based scheme and a lower bound-based scheme. 
At last, we conduct extensive experiments on four real-world datasets to demonstrate the effectiveness and efficiency of our proposed algorithms, which achieve at least \textbf{$\times$3} times faster than the baseline solution.

\end{abstract}

\ccsdesc[500]{Do Not Use This Code~Generate the Correct Terms for Your Paper}
\ccsdesc[300]{Do Not Use This Code~Generate the Correct Terms for Your Paper}
\ccsdesc{Do Not Use This Code~Generate the Correct Terms for Your Paper}
\ccsdesc[100]{Do Not Use This Code~Generate the Correct Terms for Your Paper}



\received{20 February 2007}
\received[revised]{12 March 2009}
\received[accepted]{5 June 2009}

\maketitle

\section{Introduction}
\begin{figure}[t] 
\centering

\includegraphics[scale=0.55]{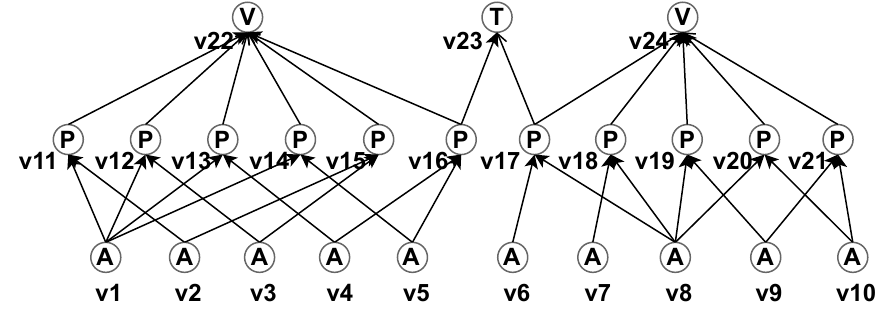}

\caption{An example of HIN}
\label{int:exampleofHIN}
\end{figure}

Heterogeneous information networks~\cite{fang2020effective} (HINs) are networks involving interconnected objects with different types
(aka \emph{labels}). Compared to the traditional homogeneous network with the same type, HIN enables to storing more rich semantic information and has become prevalent in various domains, such as citation networks \cite{zhao2023distributed}, social networks \cite{li2017most} and human-resource networks \cite{zhao2020efficient}.
Fig.~\ref{int:exampleofHIN} illustrates an HIN modeling a Database System and Logic Programming (DBLP) network, which contains four types of vertices: \emph{author} (A), \emph{paper} (P), \emph{venue} (V), and \emph{topic} (T).
The edges between different types of vertices have different semantic relationships, such as authorship (A-P) and publication (P-V).

In recent years, community search over HIN has attracted much attention due to its importance in many applications, such as recommendation~\cite{fu2020fairness, kamishima2012enhancement}, team formation~\cite{DBLP:journals/complexity/ZhaoCCL20} and identification of protein functions~\cite{DBLP:conf/ismb/DittrichKRDM08}.
Existing works like~\cite{fang2020effective, 9101354, DBLP:journals/pvldb/JiangFMCL22,DBLP:journals/pvldb/JianWC20,hu2019discovering} extend the traditional cohesive community models such as $k$-core~\cite{batagelj2003m}, $k$-truss~\cite{zhao2012large} and $k$-clique~\cite{DBLP:journals/siamdm/Feige04}, and these works require users to customize query requests like meta-path~\cite{fang2020effective, 9101354, DBLP:journals/pvldb/JiangFMCL22}, relational constraints~\cite{DBLP:journals/pvldb/JianWC20} and motif~\cite{hu2019discovering}.
However, they did not fully consider the rich semantics of the HIN, which can not handle the complex customized query request. The other important factor in the community search problem is the fairness.
As revealed in~\cite{fu2020fairness,kang2020inform,verma2018fairness}, the notion of fairness was proposed to mitigate the bias and systematic discrimination for disadvantaged people in terms of sensitive features (e.g., gender, age and race) in communities. For example, assume that people are almost composed of a specific gender in a community. In this case, members of the community are generally inclined to communicate with people of a specific gender, which causes discrimination for people of the other gender. To mitigate the discrimination in the community search problem,
Matth et al.~\cite{kleindessner2019guarantees} considered the notion of fairness as the difference in the proportion of vertex types between communities and proposed a fairness-based clustering method, which guarantees the similarity proportion of vertex type in each cluster. Similarly, Zhang et al.~\cite{zhang2023fairness} considered fairness as the difference in vertices quantity between different types of vertex in a community. However, these works only focus on group-based fairness, i.e., keeping the similarity of certain metrics between groups, which can not handle the fairness at a fine granularity level, e.g., mitigating the discrimination between members (called individual fairness) in a group.

\begin{figure}[t] 
\centering
\vspace{-5pt}
\includegraphics[scale=0.55]{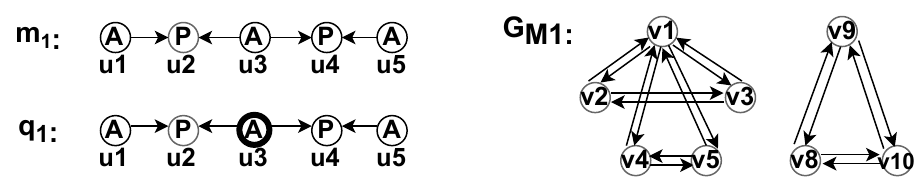}
\caption{An example of fairness community search}
\label{fig:m-graph}
\end{figure}

\textbf{IFCS problem.} In this paper, we study the problem of \textbf{I}ndividual \textbf{F}airest \textbf{C}ommunity \textbf{S}earch over HINs (IFCS) to find a set of vertices with the target type hold (1) each vertex satisfies the customized query request and has close relationships; (2) the vertices have small difference of level of activities. To formulate the IFCS problem, we face two key questions: (1) How to model the customized query requests and relationships for vertices in a community? (2) How to measure the active level of a member and the difference of active level among members in a community? 

For the former question, we extend the well-known concept of motif~\cite{hu2019discovering} to model the customized query request and relationship. A motif, also known as a \emph{higher-order structure} or \emph{graphlet}, 
is a small subgraph pattern. For example, Figure~\ref{fig:m-graph} shows two motifs $m_1, m_2$. To specify the target type, we select a vertex with user preference in a motif and use its type as the target type, and call such motif the \textit{target-aware motif}. 
For instance, Figure~\ref{fig:m-graph} illustrates a target-aware motif $q_1$, where the vertex $u_3$ with a wider border is the selected vertex. It describes that each author should collaborate on a paper with the other two authors in the community, respectively. In this case, the author $v_1$ in Figure~\ref{int:exampleofHIN} has relationships with author $v_2$ and $v_3$ through the instance of target-aware motif $\{v_3, v_{12}, v_1, v_{11}, v_2\}$.
 
For the latter question, we follow the idea in~\cite{fu2020fairness} that the active level of a member is related to the number of motif instances around it. 
In addition, we adopt the concept of the Gini coefficient~\cite{gini1921measurement} to measure the difference in active levels among members of a community. As the difference in active level among members in a community becomes small, the gini coefficient will approach zero. By carefully considering the above problems, given an HIN and a target-aware motif with a target type, the objective of IFCS is to find the community that contains vertices of the target type connected via instances of the target-aware motif and has the lowest Gini coefficient. Example~\ref{exp:1} describes a scenario where the fairest community can be found.


\begin{example}\label{exp:1}
Assume an institute wants to recruit a group of researchers for a development position such that each researcher must collaborate on at least a paper with the other two researchers of the group, respectively. This customized query request can be modeled as the target-aware motif $q_1$ in Figure~\ref{fig:m-graph}. Intuitively, if a researcher collaborates more papers with more researchers, his/her level of activity should be higher. In this case, highly active researchers in the group may be more suitable for better treatment because they have more project experience and the capability to do more work. In addition, the existence of low-activity researchers is unfair to others, because they did fewer projects but have the same treatment as others. From a fairness perspective, a group is better if researchers in the group have a small difference in level of activity. Our proposed fairest community search problem can support such a scenario.

In this case, we enumerate the instances of motif $q_1$ from the HIN in Figure~\ref{int:exampleofHIN}, and get two groups $T_{c1}=\{ v_1, v_2, v_3, v_4, v_5 \}$, $T_{c2}=\{v_8, v_9, v_{10} \}$ shown in Figure~\ref{fig:m-graph} $G_{M_1}$.
We can see that all the authors in $T_{c2}$ collaborate on a paper with the other two authors, but the authors $v_2,v_3,v_4,v_5$ in $T_{c1}$ collaborate a paper with other two authors and the author $v_1$ in $T_{c1}$ collaborate four papers with other five authors. Obviously, the activity level of $v_1$ in $T_{c_1}$ is different from others, and the activity levels of members in $T_{c_2}$ are the same. So $T_{c_2}$ have a small difference in activity level and we can identify the community $T_{c2}$ as the fairest community.

\end{example}

\textbf{Challenges.}
To find the fairest communities, a basic method is to enumerate all the communities and calculate the fairness score of each community, then return the maximal community has the lowest fairness score. However, there are two computational challenges of the basic method: (1) we need to re-enumerate an instance of motif around each vertex of the target type to verify the satisfaction of the customized query request once a vertex of the target type is removed; (2) we need to enumerate all motif instances around each member of the community to calculate their active level.

To conquer the first challenge, we propose an exploration-based filter strategy to reduce the potential target vertices that need to be checked and a message-passing based optimization strategy to avoid redundant computation. To solve the second challenge, we derive the lower bound of the fairness score to prune the unfair communities in advance. 

\textbf{Contributions.} 
We state our main contributions as follows:
\begin{itemize}


\item To the best of our knowledge, this is the first work to formalize the problem of \emph{individual fairest community search} (IFCS) over HINs, which introduces individual fairness for the community model.




\item As the IFCS problem is NP-hardness, we develop a \textit{Filter-Verify} algorithm to solve the IFCS problem.


\item We further propose an exploration-based and a message-passing based optimization strategy to reduce redundant computation, then we provide a lower bound based optimization strategy to identify and prune the unfair community in advance during the process of community search.


\item We conducted extensive experiments on four real-world datasets to demonstrate the effectiveness of the proposed fair community model and the efficiency of the proposed optimization strategies. 

\end{itemize}

The rest of this paper is organized as follows. 
First, we present 
the definition of the IFCS problem in Section~\ref{sec:problemstatement}. 
Then we describe the detailed procedure for the baseline solution of the problem in Section~\ref{sec:baseline}.
Next, we discuss the techniques in Section~\ref{sec:optimization} to speed up finding the fairest communities.
In Section~\ref{sec:experiment}, we report the experimental setting and the evaluation results. 
Finally, we review the related work in Section~\ref{sec:relatedwork}, and conclude our work in Section~\ref{sec:conclusion}.
\section{Problem Definition}\label{sec:problemstatement}


\subsection{Preliminaries}
We model an HIN as a directed graph $G=(V_G, E_G)$ with a vertex-type mapping function $\psi: V_G \rightarrow \mathcal{A}$
, where each vertex $v \in V_G$ has 
a vertex type $\psi(v) \in \mathcal{A}$. 
We use $|\cdot|$ to denote the number of elements in a set, for example, $|V_G|$ and $|E_G|$ denote the number of vertices and edges in $G$, respectively.
Here, we summarize the most important notations in Table~\ref{tab:notion}.

\begin{table}[t]
    \caption{The summary of notations}
    \vspace{5pt}
    \centering
    \begin{tabular}{p{2cm}|p{6cm}}
        \hline \textbf{Notation} & \textbf{Description} \\
        \hline $G=(V_G, E_G)$ & an HIN with vertex set $V_G$ and edge set $E_G$\\
        \hline $\psi_G(v)$
        & the type of vertex $v$ 
        in HIN $G$ \\
        \hline $\mathcal{A}$
        & the vertex type set 
        in HIN \\
        \hline $q=(V_q, E_q, v_t)$ & a target-aware motif with vertex set $V_q$, edge set $E_q$ and target vertex $v_t$  \\
        \hline $\xi_q$ & the bijective function of motif $q$\\

        \hline $g_m^q
        $ 
        & an instance of motif $q$ 
        \\
        \hline $p(v)$ & the active level of target vertex instance $v$\\
        \hline $T_c, C$ & the target-aware community and the candidate target-aware community  \\
        \hline $\operatorname{FS(T_c)}$ & the fairness score of target-aware community $T_c$ \\
        \hline
    \end{tabular}
    \label{tab:notion}
\end{table}

We first introduce the notion of motif, which has been widely used to describe subgraph patterns.

\begin{definition}[Motif~\cite{milo2002network}]\label{def:motify}

Given an HIN with its vertex-type set $\mathcal{A}$ 
, a motif $q = (V_q, E_q)$ is a small connected HIN with vertex-type mapping functions $\psi_q: V_q \rightarrow \mathcal{A}_q$,
 where $\mathcal{A}_q \subseteq \mathcal{A}$.

\end{definition}


\begin{definition}[Instance of Motif~\cite{milo2002network}]\label{def:motifyinstance}
Given an HIN $G$ with its vertex-type mapping functions $\psi$ and a motif $q=(V_q, E_q)$ with its vertex-type mapping functions $\psi_q$, a subgraph $g_m^q=(V_m^q, E_m^q)$ of $G$ is an instance of motif $q$, if $\exists$ bijection function $\xi_q: V_q \rightarrow V_m$ satisfies (1) $\forall v \in V_q$, $\psi_q(v) = \psi(\xi_q(v))$ holds;
(2) $\forall (v, v') \in E_q$, $(\xi_q(v), \xi_q(v')) \in E_m$ holds.
\end{definition}

\subsection{Target-aware Community}
To model the customized query request and formulate the target-aware community, we introduce the following new concepts.

\begin{definition}[Target-aware Motif]\label{def:querymotifanchored}
Given a motif $q = (V_q, E_q)$ with its vertex-type mapping function $\psi_q$, we can specify a vertex $v_t \in  V_q$ as 
the \emph{target vertex}.
We also use $q=(V_q, E_q, v_t)$ to represent and call it \emph{target-aware motif}.

\end{definition}  

For simplicity, all the motifs mentioned in the following are target-aware motifs.


\begin{definition}[Instances of Target Vertex] \label{targetvertexinstanceofhin}
Given an HIN $G$, a motif $q=(V_q, E_q, v_t)$ with its bijection function $\xi_q$ and the instances of motif $q$ in $G$, the instances of target vertex $v_t$ are the corresponding vertex $\xi_q(v_t)$ of the instances of motif $q$.
\end{definition}

\begin{definition}[Instances of Motif around Instance of Target Vertex] \label{targetvertexinstanceofhin}
Given an HIN $G$, a motif $q=(V_q, E_q, v_t)$ with its bijection function $\xi_q$ and an instance $v$ of target vertex $v_t$ in $G$, the instances of motif around $v$ are the instances of motif in $G$ whose corresponding vertex $\xi_q(v_t)$ = $v$.
\end{definition}


Here, we regard the instance of target vertex $v$ as \textit{$M$-neighbor} of the instance of target vertex $u$ if there exists an instance of motif around $u$ containing $v$.
In addition, we regard these two instances of target vertex $v, u$ as \textit{$M$-connected} if there exists a chain of
vertices from $v$ to $u$ such that one vertex of any two adjacent vertices in the chain is a $M$-neighbor of the other vertex.



\begin{definition}[Active Level of Target Vertex Instance]\label{def:vertexpreferencescore}
Given an HIN $G$, a motif $q = (V_q, E_q, v_t)$ and a set of $M$-connected target vertex instances $I$, the active level $s_v$ of a target vertex instance $v \in I$ is the number of motif instances $g^q_m=(V^q_m, E^q_m)$ which satisfies $\forall v' \in V^q_m \wedge \psi(v')=\psi_q(v_t)$, $v' \in I$ around $v$ in $G$.
\end{definition}

\begin{definition}[Target-aware Community]
Given an HIN $G$
with its vertex-type mapping function $\psi$
and a motif $q=(V_q, E_q, v_t)$ with its bijection function $\xi_q$, the target-aware community $T_c$ is a set of $M$-connected target vertex instances in $G$ that satisfies $\forall v \in T_c$, $s_v \geqslant 1$.
\label{def:7}
\end{definition}

Based on definition~\ref{def:7}, we call a target-aware community is maximal if it is not contained in any other target-aware community. For a motif $q$, we can also induce a homogeneous graph $G_M$ (defined in the following) from an HIN $G$, called $M$-graph, to record the $M$-neighbors of each member of the target-aware community. Essentially, the maximal target-aware communities are the vertices of the weakly connected subgraphs of $G_M$.

\begin{definition}[$M$-graph]
Given an HIN $G$ and a motif $q$, the $M$-graph is a directed homogeneous network $G_M=(V_M, E_M)$ such that (1) it contains all the M-connected target vertex instances whose active level is not less than 1; (2) for each vertex $v_M \in V_M$, it has an edge linked to each of its $M$-neighbors.

\end{definition}

\subsection{Problem Statement}

In this paper, we invoke a widely-accepted fairness measurement, i.e., Gini coefficient~\cite{gini1921measurement} 
to measure the similarity of active level among members in a target-aware community. It can be defined as follows.

\begin{definition} [Fairness Score of Target-Aware Community]\label{def:fairnessscore}
Given a target-aware community $T_c$ and the list of active levels $S$ of each member in $T_c$, the fairness score of $T_c$ can be measured as follows: 

\begin{equation}\label{equ:fairnessscore}
FS= \frac{\sum_{i=1}^{|S|}\sum_{j=1}^{|S|}\left|s_i - s_j\right|}{2 |S| \sum_{m=1}^{|S|} s_m}
\end{equation}
\end{definition}

where $s_i, s_j, s_m$ are active levels in $S$. Based on the property of Gini coefficient, the fairness score is in $(0, 1]$, where 0 represents perfect equality while 1 means maximal inequality. Intuitively, it is directly proportional to the difference of active levels among instances of target vertex and inversely proportional to the number of target vertex instance.

\begin{example}
Consider the HIN $G$ in Figure~\ref{int:exampleofHIN} and the motif $q_1$ in Figure~\ref{fig:m-graph}. We can get the M-graph $G_{M1}$ and two maximal target-aware communities $T_{c1}=\{ v_1, v_2, v_3, v_4, v_5 \}$, $T_{c2}=\{v_8, v_9, v_{10} \}$ shown in Figure~\ref{fig:m-graph}. For $T_{c1}$, the active levels of $v_1, v_2, v_3, v_4, v_5$ are 12, 2, 2, 2, 2, respectively, and the fairness score of $T_{c1}$ is 0.4. For $T_{c2}$, the active levels of $v_8, v_9, v_{10}$ is 2, 2, 2 and the fairness score of  $T_{c2}$ is 0.
\end{example}

Based on the above definitions, the target of this work is to find the maximal target-aware communities owning the minimum fairness score. Next, we formalize the problem as \underline{I}ndividual \underline{F}airest \underline{C}ommunity \underline{S}earch (IFCS) over HIN.

\begin{problem}[Individual Fairest Community Search]
Given an HIN $G = (V_G, E_G)$ and a motif $q=(V_q, E_q, v_t)$, the problem of Individual Fairest Community Search (IFCS) over HIN is to find a maximal target-aware community $T_c$ from $G$ satisfying:

\begin{equation*}
\arg \min FS(T_c) 
\end{equation*}



\end{problem}

\noindent\underline{\textbf{Problem Complexity:}} Now, we demonstrate that the IFCS problem is NP-hard by reducing the motif instance enumeration problem to it. Given an HIN $G$ that has a vertex of type $b$ and all other vertices with type $a$, we want to find a target-aware community that each vertex of type $b$ in the community should connect at least two vertices of type $a$, that is, the motif $q$ is a small HIN graph that contains a vertex with type $b$ connected with two vertices of type $a$. This is equivalent to enumerating all instances of $q$ from $G$ because the active level of each community member is calculated by enumerating all instances of $q$ in $G$. Apparently, the feasible solution of IFCS corresponds to the motif instance enumeration problem. As discussed in \cite{hartmanis1982computers}, enumerating the instance of motif from a graph is an NP-complete problem. Therefore, the IFCS problem is also an NP-hard problem.

\section{The Filter-verify Solution}\label{sec:baseline}
To address the IFCS problem, a basic solution is to enumerate all the maximal target-aware communities and calculate their fairness scores. Then, the fairest target community can be returned by selecting the one that has the minimum fairness score. In specific, we follow the same paradigm of the filter-verify algorithm~\cite{cai2020anchored}. It consists of three steps: (1) filter the unsatisfied vertices that are not in $M$-graph; (2) build the $M$-graph and calculate the active level of each vertex of $M$-graph; (3) for each weakly connected subgraph $g_M$ in $M$-graph, calculate its fairness score, then return the vertices in $g_M$ that has the lowest fairness score. The process is presented in Algorithm~\ref{met:baselinesolution}.

We first initialize a set $N_R$ to store the edges of $M$-graph and a dictionary $D$ to store the active level of each vertex of target-aware communities (Line 1). Next, we filter the unsatisfied vertices by enumerating a motif instance around the vertices of the target type (Lines 2-7). We first initialize a list $V_N$ to store the candidate vertices of $M$-graph and add vertices with type $\psi_q(v_t)$ in $G$ to $V_N$ (Line 3). For each iteration, we enumerate a motif instance around each vertex $v$ in $V_N$ using the state-of-the-art subgraph matching algorithm~\cite{DBLP:journals/tkde/SunL22}. If there is no motif instance $g_m^q$ around $v$, we can know $v$ is not included in $M$-graph and delete it from $G$ (Lines 4-6). We repeat the above process until no vertex is removed in this iteration.

Next, we build the $M$-graph and calculate the active level of each target vertex instance (Lines 8-13). For each vertex $v$ with type $\psi_q(v_t)$ in $G$, we enumerate the rest of motif instances around $v$ using the state-of-the-art subgraph matching algorithm~\cite{DBLP:journals/tkde/SunL22}. Once a motif instance $g_m^q$ around $v$ is enumerated, we update the active level of $v$ in $D$ and add the edges that connect $v$ with other vertices of target type in $g_m^q$ to $N_R$ (Lines 8-12). 
After enumerating all the motif instances around vertices in $V_N$, we can build the $M$-graph $G_M$ using edges in $N_R$ (Line 13).  
 
Finally, we get the target-aware communities and  calculate their fairness scores (Lines 14-24). We initialize $maxC$ and $maxFS$ to store the target-aware communities and the fairness score of the target-aware communities (Line 14). Then we get the maximal target-aware communities by returning the weakly connected subgraphs of $G_M$. For each weakly connected subgraph, we calculate its fairness score  using active levels stored in $D$. Once the fairness score of a community is smaller than the existing fairest communities, we remove all communities in $maxC$ and put this community to $maxC$ and the corresponding fairness score to $maxFS$ (Lines 15-24). In the end, the fairest communities in $maxC$ are the final results.

\begin{algorithm}[t]
\scriptsize
	\caption{Basic Solution}\label{met:baselinesolution}
   \KwIn{An HIN $G=(V_G, E_G)$, a motif $q=(V_q, E_q, v_t)$ with vertex type mapping function $\psi_q$.}
    \KwOut{A list of maximal individual fairness communities $maxC$}
    
    $N_R \leftarrow \varnothing$, $D \leftarrow$ empty dictionary \;
    \Repeat{$V_N \setminus $ vertices with type $\psi_q(v_t)$ in $G$ = $\varnothing$}{
        $V_N \leftarrow$ vertices with type $\psi_q(v_t)$ in $G$ \;
        \For{$\mathbf{each}$ vertex $v \in V_N$ }{
            \If{no motif instance around $v$ found by an existing subgraph isomorphism algorithm from $G$}{
                Delete $v$ from $G$ \;
            }
        }
    }

    \For{$\mathbf{each}$ instance $g_m^q = (V_m^q, E_m^q)$ of motif $q$ around $v$ in $G$ found by an existing subgraph isomorphism algorithm }{
        $\textbf{if}$ $D[v] = \varnothing$  
        $\textbf{then}$  $D[v] \leftarrow 1$ \;
        $\textbf{else}$  $D[v] \leftarrow D[v] + 1$\;
        \For{\textbf{each} vertex $v'$ with type $\psi_q(v_t)$ in $V_m^q \setminus v$}{
            $N_R \leftarrow N_R \cup \{ (v', v) \}$
        }
    }

    $G_M=(V_M, E_M) \leftarrow$ generate graph using $N_R$ \;

    $maxC \leftarrow []$, $maxFS \leftarrow 0$ \;

    \For{$\mathbf{each}$ weakly connected subgraph $g_w = (V_{w}, E_{w})$ of $G_M$}{
        \If{$|V_{w}| \geq k$}{
            $P \leftarrow$ []\;
            
            $\textbf{for}$ $v \in V_{w}$ $\textbf{do}$
                $P$.add($D[v]$)\;
                
            $FS \leftarrow$ calculate the fairness score using active levels in $P$ \algorithmiccomment{Equation~\ref{equ:fairnessscore}} \;

            \If{$FS < maxFS$}{
                $maxC$.removeAll()\; 
                $maxC$.add($V_{w}$), $maxFS = FS$ \;
            }\ElseIf{$FS = maxFS$}{
                $maxC$.add($V_{w}$)\;
            }
        }
    }

    \textbf{Return} $maxC$ \;

\end{algorithm}

\begin{example} Take the HIN $G$ in Figure~\ref{int:exampleofHIN}, the motif $q_1=(V_q, E_q, v_t)$ in Figure~\ref{fig:m-graph} as an example. Firstly we enumerate motif instances around $v_1 - v_{10}$ and get a motif instances around $v_1-v_5,v_8-v_{10}$. So we delete the vertices $v_6, v_7$. Due to the leaving of $v_6, v_7$ may cause the $v_1-v_5, v_8-v_{10}$ do not exist a motif instance around them, we re-enumerate motif instances around $v_1-v_5, v_8-v_{10}$ and get a motif instance around them, respectively. Next, we enumerate the rest of the motif instances around $v_1-v_5, v_8-v_{10}$ and record the active levels $s_{v_1}=12, s_{v_2}=2, s_{v_3}=2, s_{v_4}=2, s_{v_5}=2, s_{v_8}= 2, s_{v_9}=2, s_{v_{10}}=2$. After that, we generate the $M$-graph $G_{M1}$ shown in Figure~\ref{fig:m-graph} and get two target-aware community $T_{c1} = \{ v_1, v_2, v_3, v_4, v_5 \}$, $T_{c2} = \{ v_8, v_9, v_{10} \}$ from $G_{M1}$, and calculate the fairness score 0.4, 0 of $T_{c1}$ and $T_{c2}$, respectively. Finally, we can conclude that the fairest target-aware community is $T_{c2}$.

\end{example}

\textbf{Complexity Analysis.} The time complexity analysis of Algorithm~\ref{met:baselinesolution} consists of the following steps. We first construct the index discussed in~\cite{DBLP:journals/tkde/SunL22} to enumerate instances of motif, which takes $O(|E_G| \cdot |E_q|)$ time. For each iteration of target vertex identification process, it takes $O(d^{\frac{3}{2}d_t} \cdot |V_t|)$ to enumerate a motif instance for each vertex in $V_t$ using the algorithm in~\cite{DBLP:conf/sigmod/HanKGPH19}, where $d$ is the average degree of HIN $G$, $d_t$ is the length of the shortest path between target vertex and its most distanced node in motif $q$, and $|V_t|$ is the number of vertices of target type. In the worst case, the number of iteration could be $|V_t|$, so the total time complexity of the target vertex identification process is $O(d^{\frac{3}{2}d_t} \cdot V_t^2)$. In the process of active level calculation, the time complexity is $|V_t|^{|V_q|}$ because the number of motif instances
could be $|V_t|^{|V_q|}$. In the process of maximal target-aware communities generation, the time cost is $O(|V_t| + |E_G|)$ by returning the weakly connected subgraphs of $G_M$.
Thus, the total time complexity of Algorithm~\ref{met:baselinesolution} is $O(d^{\frac{3}{2}d_t} \cdot V_t^2 + |V_t|^{|V_q|} + |E_G| \cdot |E_q|)$ in total.

It is obvious that the complexity of basic solution is high. The main drawbacks lie in: (1) all the vertices of target type need to be identified for each iteration in target vertex identification process; (2) all the vertices in $M$-graph need to enumerate the motif instances around them to calculate their active level.

\section{Optimization}\label{sec:optimization}
To overcome the above drawbacks, in this section, we first propose an exploration-based filter to prune the ineligible vertices of the target type. Then, we develop a message-passing based optimization strategy to avoid redundant computation. Finally, we propose a lower bound-based to filter the unfair community in advance by using the derived lower bound of fairness score.

\subsection{Reducing Potential Target Vertices}
In this subsection, we propose an exploration-based filter to further reduce potential instances of target vertex. Before introducing the details, we first introduce a query vertex filtering strategy, called Neighborhood Label Frequency (NLF) filter~\cite{bi2016efficient}. It aims to find the candidate vertices of a query vertex in a motif that may be contained in instances of motif. 

\begin{definition}[Neighborhood Label Frequency (NLF) Filter~\cite{bi2016efficient}]\label{lem:NLF}
Given a query vertex $u$ in a motif $q$ and a vertex $v$ in an HIN $G$, $v$ is the candidate vertex of $u$ if $v$ satisfies $\forall l \in L_N(u)$, $d_i(v,l) < d_i(u, l) \land d_o(v,l) < d_o(u, l)$.
\end{definition}

where $L_N(v)$ is the set of unique labels of $u$'s in-neighbors and out-neighbors, $d_i(v,l)$ is the number of in-neighbors of $v$ with label $l$, and $d_o(v,l)$ is the number of out-neighbors of $v$ with label $l$. In addition, we also introduce the other candidate vertex filter method, which supports our exploration-based filter search.

\begin{definition}[Exact Star Isomorphism Constraint~\cite{DBLP:journals/tkde/SunL22}]\label{def:advfilter}
    Given an HIN $G$, a query vertex $u$ of a motif $q$ and a candidate vertex $v$ of $u$ pass the NLF filter in $G$, $v$ satisfies the exact star isomorphism constraint if $\forall u' \in N(u)$, $\exists v' \in N(v)$ such that $v' \in u'.C$.
\end{definition}

where $N(u)$ is the in-neighbors and out-neighbors of $u$, $u'.C$ is a set of candidate vertices of $u'$ in HIN $G$. Intuitively, if the candidate vertex $v$ of $u$ satisfies the exact star isomorphism constraint, there exists at least one candidate vertex of $u'$ in neighbors of $v$ for each neighbor $u'$ of $u$. Here, we adopt the breadth-first search (BFS) order starting from the vertex $v_{t}$ in $q$ to find the candidate vertices of each query vertex. Based on the exact star isomorphism constraint and BFS order, we have the following corollary.


\begin{corollary}\label{cor:advfilter}
    Given an HIN $G$, a query vertex $u$ of a motif $q$, the BFS order $\pi$ of $q$ and a vertex $v$ in $G$. If $v$ satisfies the exact star isomorphism constraint, it must hold the following two conditions: (1) $\forall u' \in N(u) \wedge idx_{u'}(\pi) < idx_u(\pi) $, $\exists v' \in N(v)$ such that $ v' \in u'.C $; (2) $\forall u' \in N(u) \wedge  idx_{u'}(\pi) > idx_u(\pi) $, $\exists v' \in N(v)$ such that $ v' \in u'.C $; 
\end{corollary}
where $idx_u(\pi)$ is the position (i.e., index) of $u$ in the searching order $\pi$. Based on the NLF filter and corollary~\ref{cor:advfilter}, we can find the candidate target vertex instances and explore the candidate regions around each candidate target vertex instance that may contain motif instances around it. Intuitively, the candidate region is composed of the candidate vertices of each query vertex. In this case, the candidate $M$-neighbors of a candidate target vertex instance $v$ are the vertices of the target type in the candidate region. We use a directed homogeneous network, denoted as $CM$-graph, to store the candidate $M$-connected target vertex instances.

\begin{definition}[$CM$-graph]
\label{def:follower}
Given a HIN $G=(V_G, E_G)$ and a motif $q=(V_q, E_q, v_t)$, the $CM$-Graph is a directed homogeneous graph $G_{CM}=(V_{CM}, E_{CM})$ such that (1) it contains all the vertices of target type $v_{CM} \in V_{CM}$ passing the NLF filter and satisfying the constraints in corollary~\ref{cor:advfilter}; (2) for each vertex $v_{CM} \in V_{CM}$, it has an edge linked to each of its candidate $M$-neighbors.
\end{definition} 
 
We process the exploration-based filter in two steps: (1) generate the candidate regions and $CM$-graph to explore the candidate $M$-connected vertices using condition 1 of Corollary~\ref{cor:advfilter} in the forward candidate exploration; (2) refine the candidate regions and $CM$-graph to prune ineligible candidate $M$-connected vertices using condition 2 of Corollary~\ref{cor:advfilter} in the backward candidate refinement; 

\noindent\textbf{Forward Candidate Exploration}. We first initialize $S$ to store the edges of candidate regions, $E_{CM}$ to store the edges in $CM$-graph, and $\pi$ to store the BFS order of motif $q$ (Line 1). Then we get the candidate target vertex instances $C$ from HIN $G$ by selecting the vertices of the target type passing the NLF filter (Lines 2-3), and delete the vertices of type $\psi_q(v_t)$ in $V_G$ but not contained in $C$ to reduce the searching space (Line 4). Next, we explore the candidate region around each candidate target vertex instance $c \in C$ following the BFS-order $\pi$ in the forward candidate exploration process (Lines 5-31). 

Specifically, we initialize $S'$ to store the set of edges in the candidate region around $c$, u.$C'$ to store the candidate vertices of a query vertex $u$, and use a boolean variable i-add to denote whether $v$ has a candidate region around it (Line 6). Intuitively, i-add is false if a query vertex has no candidate vertex, which means there is no candidate region around $v$. Then we get the candidate vertex of each vertex $u \in \pi \setminus v_t$ in the candidate region around $c$ (Lines 7-22). 
We first get the neighborhoods $N'_u$ of $u$ whose candidate vertices have been found (Line 9). Then we randomly select a vertex $u_b$ in $N'_u$ and use its candidate vertices to find the candidate vertex of $u$ (Lines 10-22). To achieve this, for each candidate vertex $v$ of $u_b$, we first find the neighbours $N(v)$ of $v$ satisfying: (1) have the same type as $u$; (2) pass the NLF filter, and (3) have the same edges direction $dir(v, v')$ between $v$ and $v' \in N(v)$ as edges direction $dir(u_b, u)$ between vertices $u_b$ and $u$ (Line 12). If there does not exist such neighbor, the i-add is set to be false (Line 22). Otherwise, we verify whether $v'$ satisfies condition 1 of Corollary~\ref{cor:advfilter} (Lines 13-21). 

We use $E_t$ to store the edges between $v'$ and its neighbors that are contained in the candidate vertices of query vertices $N'_u \setminus u_b$ (Line 13), and use a boolean variable c-add to denote whether $v'$ satisfies the condition 1 of Corollary~\ref{cor:advfilter} (Line 14). For each query vertex $\hat{u} \in N'_u \setminus u_b$, we add the edges between candidate vertices of $\hat{u}$ and $v'$ into $E'_t$ (Line 16). If $E'_t$ is empty, we know that there is no candidate vertex of $\hat{u}$ around $v'$, i.e., $v'$ does not satisfy the condition 1 of Corollary~\ref{cor:advfilter} (Lines 17-18). Otherwise, we add edges in $E'_t$ to $E_t$ (Line 19). If $v'$ is eligible, we add edges in $E_t$ to $S'$ (Lines 20-21). Once $v$ is an eligible candidate vertex instance, we add the edges in $S'$ to $S$ and add candidate vertices of query vertex $u$ in motif to the candidate target vertex set $\hat{u}.C$ (Lines 23-24). Finally, we get the candidate $M$-neighbors of $c$ by selecting the vertices with type $\psi_q(v_t)$ in the candidate region around $c$ except $c$, then add edges between $c$ and its candidate $M$-neighbors into $E_{CM}$ (Lines 25-26). After that, we filter the ineligible $M$-connected candidate target vertex instance whose quality is less than $k$ (Lines 27-29) and delete the vertices with type $\psi_q(v_t)$ in $V_G$ but not included in $v_t.C$ to reduce the searching space in backward exploration process (Lines 30-31).

\noindent\textbf{Backward Candidate Refinement.} In the backward processing, we refine each candidate region based on the unexploited neighbors of each query vertex $u$ in the forward candidate exploration process, i.e., the in-neighbors and out-neighbors $N(u)$ of $u$ whose index is larger than $u$ in BFS order (Lines 32-41). In contrast to the forward processing, now we find the candidate vertices of each query vertex following the in reverse order of $\pi$ and get the refined candidate region of each vertex of target type in $u_f.C$.

\begin{algorithm}[t]
\scriptsize
    \caption{Exploration-based Filter$(G,q)$}\label{met:explorationbasedfilteringsearch}
    
    \KwIn{An HIN $G=(V_G, E_G)$, a motif $q=(V_q, E_q, v_t)$ with vertex type mapping function $\psi_q$}
    \KwOut{a $CM$-Graph and a refined $G$}

        $S \leftarrow \varnothing$, $E_{CM} \leftarrow \varnothing$, $C \leftarrow \varnothing$, $\pi \leftarrow$ BFS order of $q$  \;
        \For{$\mathbf{each}$ vertices $v \in G$}{
            $C \leftarrow C \cup \{v \mid \psi(v)=\psi_q(v_t), v$ pass the NLF filter
 $\}$ \;
        } 

        Delete vertices $\{ v \in V_G \mid \psi(v) = \psi_q(v_t) \} \setminus C$ from $G$ \;

        \tcp{\textbf{Lines 5-31: Forward candidate Exploration}}
        \For{$\mathbf{each}$ vertices $c \in C$ }{
            $S' \leftarrow \varnothing$ , $v_t.C' \leftarrow \{c\}$, i-add $\leftarrow $ True \;
            \For{$\mathbf{each}$ query vertex $ u \in \pi$  $\setminus v_t$}{
                $u.C' \leftarrow \varnothing$ \;
                $N'_u \leftarrow \{ u' \in N(u) \mid idx_u(\pi) < idx_{u'}(\pi) \}$ \;

                $u_b \leftarrow$ Random select a vertex from $N'_u$ \;

                \For{$\mathbf{each}$ vertex $v \in u_b.C'$}{
                    \For{$\mathbf{each}$ vertex $v' \in \{\hat{v} \in N(v) \mid \psi(\hat{v}) = \psi_q(u)$ $\wedge$ $v$ pass the NLF filter $\wedge$ $dir(v, \hat{v})=dir(u_b, u) \}$}{ 
                        $E_t \leftarrow$ edges between $v$ and $v'$ with the same direction as edges between $u_b$ and $u$\;
                        c-add $\leftarrow $True \;
                        \For{$\mathbf{each}$ vertex $\hat{u} \in N_u' \setminus u_b$}{
                            $E'_t \leftarrow$ $\{ (\bar{v},v') \in E(v') \setminus (v, v') | \bar{v} \in \hat{u}.C'  \}$ \; 
                            \If{$E'_t = \varnothing$}{
                                c-add $\leftarrow False$, \textbf{Break} \;
                            }$\textbf{else}$
                                $E_t \leftarrow E_t \cup E'_t$  \;
                        }
                        \If{c-add = True}{
                            $S' \leftarrow S' \cup E_t$, $u.C' \leftarrow u.C' \cup \{ v' \}$ \;
                        }
                    }
                }
                \textbf{if} $u.C' \leftarrow \varnothing$ \textbf{then} i-add $\leftarrow $ false, \textbf{Break};
            }

            \If{i-add = true}{
                $S \leftarrow S \cup S'$, \textbf{for each} $u \in \pi$  \textbf{do} $u.C \leftarrow u.C'$ \;
                $G_c=(V_c, E_c) \leftarrow$ Induce graph from $G$ using $S'$ \;
                $E_{CM} \leftarrow E_{CM} \cup \{ (c, \hat{v}) \mid \hat{v} \in V_c \setminus c \wedge \psi(\hat{v}) = \psi_q(v_t) \}$ \;
            }
        }

        $G_{CM}=(V_{CM}, E_{CM}) \leftarrow $ Induce graph using $E_{CM}$ \;
        Delete vertices not include in $v_t.C$ from $G_{CM}$ \;


        Remove the connected subgraphs whose number of vertices is smaller than $k$ from $G_{CM}$\;

        $G=(V_{G}, E_{G}) \leftarrow $ Induce graph from $G$ using $S$ \;
        Delete vertices $\{ v \in V_{G} \mid \psi(v) = \psi_q(v_t) \} \setminus V_{CM}$ from $G$ \;

        \tcp{ \textbf{Lines 32-40: Backward candidate refinement}}

        $u_f \leftarrow$ last vertex in $\pi$ \;
        \For{\textbf{each} vertices $v \in u_f.C \cap V_{G'}$}{
             $S' \leftarrow \varnothing$ , $v_f.C' \leftarrow \{c\}$, i-add $\leftarrow $ True \;
             \For{$\mathbf{each}$ query vertex $ u \in \pi$ $\setminus u_f$ in reverse order}{
                Same as Lines 8 \;
                $N'_u \leftarrow \{ u' \in N(u) \mid idx_u(\pi) > idx_{u'}(\pi) \}$ \;
                Same as Lines 10-22 \;
             }
             Same as Lines 23-24 \;
        }

        Delete vertices not include in $v_t.C$ from $G_{CM}$ \;

        Same as Line 27-31 \;

        \textbf{Return} $G_{CM}$, $G$ \; 
            
\end{algorithm}

    



    

        



\begin{figure}[t]
    \centering
    \vspace*{-6mm}
     \subfloat[An example of motif]{{\includegraphics[scale=0.68]{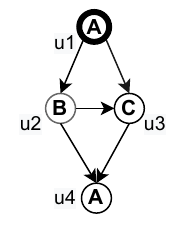}\label{fig:t-motif} }}
     \subfloat[An example of HIN]{{\includegraphics[scale=0.68]{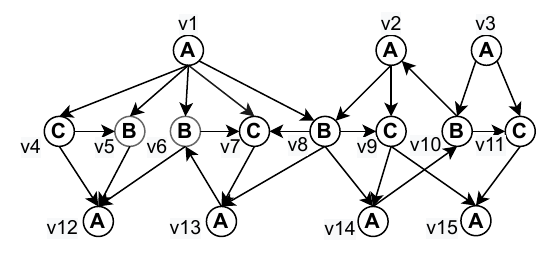}\label{fig:datagraphoftmotif} }}

    \caption{Example of exploration-based filtering search}
    \label{fig:my_label}
\end{figure}

\begin{example}
Consider the motif $q$ in Figure~\ref{fig:t-motif} and the HIN in Figure~\ref{fig:datagraphoftmotif}. Firstly, we get the BFS order $\{u_1, u_2, u_3, u_4\}$ of $q$ and the candidate target vertex instances $C=\{ v_1, v_2, v_3 \}$. Next we explore the candidate region around each candidate target vertex instance in $C$. 

For $v_1$, (1) in the forward processing, we first find the candidate vertex of $u_2$ with $N'_u = \{ u_1 \}$ and $u_b = u_1$. Then we process each vertex $v'$ in $\{v_6, v_8\}$. Note that although $v_5$ is the neighbor of $v_1$ with the same label as $u_2$, it is pruned because it does not pass the NLF filter. Due to the $N'_u \setminus u_b$ is empty, we conduct that $v_6, v_8$ are candidate vertex of $u_2$ and add the edges $(v_1, v_6), (v_1, v_8)$ in $E_t$ to the $S'$. Next, we find that $v_7, v_9$ are the candidate vertex of $u_2$, but $v_9$ will be pruned because it does not satisfy the condition 1 of Corollary~\ref{cor:advfilter}, i.e., there is no candidate vertex of $u_1$ in the neighbors of $v_9$. Then we add $(v_6, v_7), (v_8, v_7), (v_1, v_7)$ into $S'$. Finally we find $v_{13}$ as the candidate vertex of $u_4$ and add $(v_{13}, v_7), (v_{13}, v_8)$ into $S'$. The candidate region around $v_1$ is the subgraph induced by edges $S' = \{ (v_1, v_6), (v_1, v_8), (v_6, v_7), (v_8, v_7), (v_1, v_7), (v_{13}, v_7), (v_{13}, v_8)\}$ and the candidate vertex $C'$ of each query vertex is $u_1 = \{ v_1 \}$, $u_2 = \{ v_6, v_8 \}$, $u_3 = \{ v_7 \}$ and $u_4 = \{ v_{13} \}$. 

(2) In the backward processing, we first process the candidate vertex of $u_3$, i.e., $v_7$, with $N'_u = \{u_4\}$ and $u_b = u_4$. Then we find the edges $\{ (v_{13}, v_7) \}$ connected with $v_7$ and candidate vertices of $u_b$ i.e., \{ $v_{13}$ \}, and add them to $E_t'$. Next we process the $u_2$'s candidate vertices $\{ v_6, v_8 \}$. We pruned $v_6$ because it does not pass the NLF filter, so we add edge $\{(v_{13}, v_8), (v_8, v_7)\}$ into $E_t'$. Finally we process the $u_1$'s candidate $v_1$ and add the edges $\{ (v_1, v_7), (v_1, v_8) \}$ to $E'$. The candidate region around $v_1$ after backward refinement is the subgraph induced by edges $S' = \{ (v_1, v_7), (v_1, v_8), (v_8, v_7), (v_{13}, v_7), (v_{13}, v_8)\}$.

\end{example}

The time complexity of Algorithm~\ref{met:explorationbasedfilteringsearch} consists of the following steps. In the forward candidate exploration process, we need to take $O(|E_G| \cdot d_q(u))$ to check the candidate vertices around query vertex $u$, where $d_q(u)$ is the degree of vertex $u$ in motif $q$. Thus the total running time of the forward candidate exploration process is $O(\sum_{u \in V_q}|E_G| \cdot d_q(u))$, i.e., $O(|E_G| \cdot |E_q|)$. In backward candidate refinement, we have the same process to check the candidate vertices around each query vertex $u$ in motif $q$. So the total complexity of Algorithm~\ref{met:explorationbasedfilteringsearch} is $O(|E_G| \cdot |E_q|)$.


 
\subsection{Message-passing based Optimization Strategy}\label{sec:4.2}

\begin{figure}[t] 
\centering

 \subfloat[An example of HIN]{{\includegraphics[scale=0.65]{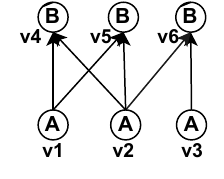}\label{int:datagraph4.2} }} \quad \quad \quad
\subfloat[An example of Motif]{{\includegraphics[scale=0.65]{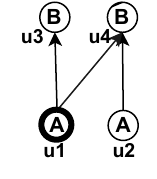}\label{int:tmotif4.2} }}

\caption{An example of message-passing strategy}

\end{figure}

By reviewing the process of filter-verify solution, we can see that all the vertices of the target type need to re-enumerate a motif instance around them in each iteration to check whether they are in $M$-graph. For example, considering the motif in Figure~\ref{int:tmotif4.2} and the HIN in Figure~\ref{int:datagraph4.2}, if we remove $v_3$ by NLF filter, we need to re-enumerate a motif around $v_1, v_2$ in the next iteration. However, $v_1$ do not need to re-enumerate after removing $v_3$ because the deleting of $v_3$ does not affect the structure around $v_1$. To solve this challenge, we propose a message-passing based strategy to avoid unnecessary re-enumeration. 

The key idea of our strategy is that the target vertex instance does not need to re-enumerate a motif around it if none of its $M$-neighbors is removed.
Recall that the candidate target vertex instances and their candidate $M$-neighbors have been stored in the $CM$-graph. Thus, if a candidate target vertex instance is removed from $CM$-graph, we can send a message to its in-neighbors in $CM$-graph to re-identify whether there exists a motif instance around them in the next iteration.

The detailed procedure is presented in Algorithm~\ref{met:messagepassingbasedsearch}. We first initialize $V_e$ to store the candidate vertex instances that need to check whether it has a motif instance around it (Line 1), then we iteratively check the instances in $V_e$ until $V_e$ is empty (Lines 2-8). We first initialize $V'_e$ to record the candidate target vertices whose candidate $M$-neighbors are removed (Line 3). For each vertex instance $v_e$ in $V_e$, once a motif instance around $v_e$ is enumerated by an existing subgraph isomorphism, we immediately check the next vertex instance. Otherwise, we add the in-neighbors of $v_e$ into $V'_e$ and delete $v_e$ from $G_{CM}$ and $G$ (Lines 4-7). Once the checking process is finished, if $V'_e$ is not empty, we continue to check the candidate vertex in $V'_e$ in the next iteration (Line 8).
Finally, we return the HIN $G$ (Lines 9).


\begin{algorithm}[t]
\scriptsize
	\caption{Message-Passing$(G, q, G_{CM})$}\label{met:messagepassingbasedsearch}

	\KwIn{An HIN $G=(V_G, E_G)$, a motif $q=(V_q, E_q, v_t)$ with vertex type mapping function $\psi_q$ and a $CM$ graph $G_{CM}=(V_{CM}, E_{CM})$}
	\KwOut{A list of individual fairness community $maxC$}
 
    $V_e \leftarrow$ vertices with type $\psi_q(v_t)$ in $G$ \;

    \While{$V_e \neq\varnothing$ }{
        $V'_e \leftarrow \varnothing$ \; 
        \For{\textbf{each} vertex $v_e$ in $V_e$}{
            \If{ no motif instance around $v_e$ found by an existing subgraph isomorphism algorithm from $G'$ }{ 
                $V'_e \leftarrow V'_e \cup \{$in-neighbors of $v_e$ in $G_{CM}$$\}$ \;
                Delete $v_e$ from $G_{CM}$ and $G$ \;
            } 
        }
        $V_e \leftarrow V'_e$
    }
    
    \textbf{Return} $G$\; 
    
\end{algorithm}



\subsection{The Lower Bound of Fairness Score}
In this subsection, we propose a lower bound of fairness score to filter the unfair communities in advance. The key idea is that if we have calculated the active levels of some $M$-connected target vertex instances and know the lower bound of the fairness score is not lower than the fairness score of the existing fairest communities, we immediately know that the $M$-connected target vertex instances are not fairest and do not need to enumerate motif instances around the rest $M$-neighbor target vertex instances. Before providing a detailed proof of the lower bound, we provide an equivalent formula of the Equation~(\ref{equ:fairnessscore}) proved in~\cite{zagier1983inequalities}.

\begin{equation} \label{equ:fairnessscore2}
FS=\frac{2 \left(s_{1}+2 s_{2}+\ldots+|S| s_{|S|}\right)}{|S| \sum_{j=1}^{|S|} s_j} - \frac{|S|+1}{|S|}
\end{equation}

subject to

\begin{equation*}
s_1 < s_2 < \ldots < s_{|S|}
\end{equation*}

which can be transferred into the following form:

\begin{equation}\label{equ:pro}
    FS = \frac{s_{1}+3 s_{2}+\ldots+(2|S|-1) s_{|S|}}{|S|\sum_{j=1}^{|S|} s_j} - 1
\end{equation}

subject to

\begin{equation*}
s_1 < s_2 < \ldots < s_{|S|}
\end{equation*}

Based on the Equation~(\ref{equ:pro}), we can get the following property.

\begin{property}\label{pro:1}
Given a candidate target-aware community $C$ and a list of active levels $S'$ of $n$ $M$-connected target vertex instances $C'$ in $C$, the active levels $S''$ of each target vertex instance in $C \setminus C'$ should be smaller than the maximum value $S'_{max}$ of $S'$ if we intend to minimize the fairness score $FS$ of $C$.
\end{property}

\begin{proof}
We prove the correctness of this property by contradiction. 
Assume $S = S' \cup S''$ has been sorted in ascending order. Through using the Equation~\ref{equ:pro}, fairness score of $C$ can be written as $FS = \frac{ \sum_{i=1}^{|C|} (2i - 1)s_i}{|C|\sum_{i=1}^{|C|} s_i} - 1$, where $s_i \in S$. It's a multivariate equation where the elements in $S''$ are independent variables. The first-order partial derivatives of $FS$ is $\frac{\partial FS}{\partial s_p} = \frac{(2p-1)B-A}{(s_p + B)^2}$, where $s_p \in S''$, $A, B$ are constants and
 $A = \left(\sum_{j=1}^{|C|} (2j-1) s_j \right) - (2p-1)s_p$, $B=(\sum_{j=1}^{|C|} s_j)-s_p$. In this case, if $s_{|C|} > S'_{max}$, $s_{|C|}$ must in $S''$, and the first-order partial derivatives of $FS$ w.r.t $s_{|C|}$ is positive because $(2|C|-1)B > A$. Therefore, $s_{|C|}$ must be less than or equal to $S'_{max}$ if we want to minimise $FS$ because $FS$ will become larger as $s_{|C|}$ becomes larger when $s_{|C|}$ is greater than $S'_{max}$. Through the above inference, we can determine that other active levels in $S''$ must be less than or equal to $S'_{max}$ iteratively. 
\end{proof}

In addition, we proved the following property to support the lower bound of the fairness score.

\begin{property}\label{pro:2}
    Given a list of numbers $L = {l_1, \ldots, l_m}$, each number in $X = {x_1, \ldots, x_n}$ should be equal to the median value of $L$ if we intend to minimize $z = \sum_{i=1}^{|S|}\sum_{j=1}^{|S|}\left|s_i - s_j\right| $, where $S = L \cup X$ and $s_i, s_j \in S$.

\end{property}

\begin{proof}
     We rewrite $z$ to $z = (2 \sum_{j=1}^{|L|} \sum_{i=1}^{|X|} |l_j - x_i|) + \sum_{i=1}^{|X|} \sum_{j=1}^{|X|} |x_i - x_j|$. As proved in \cite{noah2007median}, $\sum_{j=1}^{|L|} |l_j-x|$ is minimal if $x$ is equal to the median of $S$. Thus, $(2 \sum_{j=1}^{|L|} \sum_{i=1}^{|X|} |l_j - x_i|)$ can be minimized when $x_i \in X$ is the median of $L$. In addition, $\sum_{j=1}^{|X|} |x_i - x_j|$ can be minimized when $x_i \in X$ are the same. Thus, we can conclude $z$ can be minimized when each number in $X$ equals the median value of $L$.
\end{proof}

Now, we show a lower bound of the fairness score in Property~\ref{pro:lowerbound} based on Property~\ref{pro:1}, \ref{pro:2} and Equation~\ref{equ:fairnessscore}.

\begin{property}\label{pro:lowerbound}
Given a candidate target-aware community $C$ and a list of active levels $S'=\{ s_1 \ldots s_m \}$ of $m$ $M$-connected target vertex instances $C'$ in $C$, the lower bound $FS^{LB}$ of the fairness score can be calculated as:

\begin{equation}\label{equ:lowerbound}
FS^{LB}=\frac{\sum_{i=1}^{|S|} \sum_{j=1}^{|S|}|s_i-s_j|}{2|C|\left(\sum_{m=1}^{|S'|}s_m  + (|S|-|S'|)S'_{max} \right)}
\end{equation}

\end{property}

where $S$ contains the active levels in $S'$ and $|C|-|S'|$ median value of $S'$, $s_i, s_j \in S$ and $s_m \in S'$.

\begin{proof} We get the lower bound of $FS$ by minimize the numerator and maximize the denominator of Equation~\ref{equ:fairnessscore} separately. Based the property~\ref{pro:2}, we can get the minimum value of numerator $\sum_{j=1}^{|S|}|s_i-s_j|$ when the active levels of target vertex instances $C \setminus C'$ equal to the medium value of $S'$. To get the maximum value of the denominator $2|S|\left(\sum_{=1}^{|S|}s_m\right)$, we need to get the maximum value of $|S|$ and $\sum_{=1}^{|S|}s_m$. Recall that the target-aware community is contained in the candidate candidate community, thus the maximum value of $|S|$ are the size of candidate community $|C|$. As discussed in Property~\ref{pro:1}, $\sum_{=1}^{|S|}s_m$ have the maximum value when the active levels of target vertex instances $C \setminus C'$ are equal to the maximum value of $S'$. Based on the maximum and minimum value of the denominator and numerator, we can get the lower bound of fairness score.
\end{proof}


\subsection{The optimization Algorithm}

Algorithm~\ref{met:lowerbound} presents the optimization algorithm by reducing potential target vertices and utilizing message-passing based strategy and the lower bound of fairness score. Initially, we remove the ineligible vertices of the target type using Algorithm~\ref{met:explorationbasedfilteringsearch} (Line 1) and find the vertices in $M$-graph using Algorithm~\ref{met:messagepassingbasedsearch} (Line 2). Then we initialize $D, maxC, FS_m$ to store active levels, the fairest communities and their fairness score (Line 3). Next, we search the target-aware community in each candidate target-aware community $g$, i.e., each weakly connected subgraph of $G_{CM}$ using the lower bound of fairness score (Lines 4-35). For each candidate target-aware community, we first initialize $Vst, IV, UV, VIV$ to store the visited $M$-connected target vertex instances, in-neighbors of visited target vertex instances, unvisited $M$-connected vertices of visited target vertex instances and visited target vertex instances but not $M$-connected by target vertex instances in $Vst$ (Line 5). Then we randomly add a vertex from $g$ to $UV$ (Line 6), and start to find the target-aware community by visiting the vertex in $UV$(Lines 7-35). For each vertex $v$ in $UV$, we add it to $Vst$, enumerate the motif instances around it to calculate its active level, and add the $M$-neighbors of $v$ which is not visited to $UV$ (Lines 8-12). Based on the active levels of visited target vertex instances, we calculate the lower bound of the fairness score. If the lower bound is higher than the fairness score of existing fairest communities, we remove this candidate target-aware community; else, we continue to implement the above process (Lines 13-15). If $UV$ is empty, we consider whether the in-neighbors $OV$ of vertices in $Vst$ but not contained in $Vst$ and the vertices in $Vst$ are $M$-connected (Lines 16-35). We first get the $M$-neighbors of each vertex $\hat{v}$ in $OV$ (Line 17). If $\hat{v}$ is contained in $VIV$, we can get the $M$-neighbors of $\hat{v}$ by collecting the out-neighbors of $\hat{v}$ in $g$ (Lines 33-35); else, we get its $M$-neighbors and active level by enumerating the motif instances around $\hat{v}$ (Lines 18-26). If $M$-neighbors of $\hat{v}$ are contained in $Vst$, we know that $\hat{v}$ is $M$-connected to the vertices in $Vst$ (Line 27-29); if not, we record the $M$-neighbors and active level of $\hat{v}$ (Lines 30-32). We repeat the above process until $UV$ and $OV$ are empty. Finally, we return the fairest communities stored in $MaxC$.

The time complexity of Algorithm~\ref{met:lowerbound} is $O(d^{\frac{3}{2}d_t} \cdot V_t^2 + |V_t|^{|V_q|} + |E_G| \cdot |E_q|)$ in total. But the exploration-based filter and message-passing based optimization strategy help to filter out ineligible vertices in $V_t$, and the lower bound based pruning rules filter the unfair communities so as to improve the efficiency.

\begin{algorithm}[t]
\scriptsize
	\caption{The Optimization Algorithm}\label{met:lowerbound}
  
	\KwIn{An HIN $G=(V_G, E_G)$, a motif $q=(V_q, E_q, v_t)$ with vertex type mapping function $\psi_q$}
	\KwOut{A list of individual fairness communities $maxC$}

        $G_{CM} \leftarrow$ Exploration-based Filter$(G, q)$ \;
        $G \leftarrow$ Message-Passing$(G, q, G_{CM})$ \;
        $D \leftarrow$ empty dictionary, $FS_m=1$, $maxC=[]$ \;

        \For{\textbf{each} weakly connected graph $g=(V_g, E_g)$ of $G_{CM}$ in ascending order of $|g|$}{
            $Vst \leftarrow \varnothing$, $i_f \leftarrow $ False, $IV \leftarrow \varnothing$, $UV \leftarrow \varnothing$ $VIV \leftarrow \varnothing$\;

        $UV \leftarrow$ Random select a vertex from $V_g \setminus Vst$ \;

        \While{$UV \neq \varnothing$}{
            Random pop a vertex $v$ from $UV$ to $Vst$ \;
            $IV \leftarrow IV \cup \{$ in-neighbors of $v$ in $g$ $\}$ \;
            \For{\textbf{each} instance $g_m^q=(V_m^q, E_m^q)$ of motif $q$ around $v$ in $G$}{
                Same as Line 9-10 of Algorithm~\ref{met:baselinesolution} \;
                $UV \leftarrow UV \cup \{ v' \in V_m^q \mid \psi(v')=\psi_q(v_t) \wedge v' \notin Vst \} $ \;
            }
            \If{$i_f =$ False}{
                $FS^{LB} \leftarrow$ calculate the lower bound using active levels of $Vst$\;
                $\textbf{if}$ $FS^{LB}$ > $FS_m$ $\textbf{then}$ $i_f \leftarrow $ True \;
                
            }
            \If{$UV = \varnothing$}{
                 $OV \leftarrow IV \setminus Vst$, $IV \leftarrow \varnothing$ \;
                 \If{$OV = \varnothing$}{
                    Same as Line 19-24 of Algorithm~\ref{met:baselinesolution} \;
                 }\Else{
                    \For{\textbf{each} vertex $\hat{v} \in OV$}{
                        \If{$\hat{v} \notin VIV$}{
                            $MN \leftarrow \varnothing$ \;
                            \For{\textbf{each} instance $g_m^q=(V_m^q, E_m^q)$ of motif $q$ around $\hat{v}$ in $G$}{
                                Same as Line 9-10 of Algorithm~\ref{met:baselinesolution} \;
                                $MN \leftarrow MN \cup \{ v' \in V_m^q | \psi(v')=\psi_q(v_t) \}$ \;
                            }
                               
                            \If{$MN \cap Vst \neq \varnothing$}{
                                $UV \leftarrow UV \cup (MN \setminus Vst)$ \;
                                $Vst \leftarrow Vst \cup \{ \hat{v} \}$ \;
                            }\Else{
                                $VIV \leftarrow VIV \cup \{ \hat{v} \}$\;
                                Delete out-edges of $\hat{v}$ in $g$ and add out-edges between $\hat{v}$ and vertices in $MN$ to $g$ \;
                            }
                        }\Else{
                            $MN \leftarrow$ out-neighbors of $\hat{v}$ in $g$ \;
                            Same as Line 27-29 \;
                        }
                    }
                 }
            }
        }
    }

        \textbf{Return} $maxC$ \;

\end{algorithm}

\section{Experiments} \label{sec:experiment}
\subsection{Experimental Setup}

\footnotetext[1]{\url{https://pytorch-geometric.readthedocs.io/en/latest/modules/datasets.html}}

\textbf{Dataset.}
We performed extensive experiments on four real-world HIN datasets: IMDB\footnotemark[1], 
DBLP\footnotemark[1],
Freebase\footnotemark[1], 
and Amazon\footnotemark[1]. Their statistics, such as the number of vertices, edges, vertex types, average degree of vertices and number of distinct motifs, are presented in Table~\ref{table:dataset}.  
IMDB is an online dataset of movies and television programs, which consists of three types of vertices (movies, directors, and actors). 
DBLP is a website for computer science bibliography, which has four vertex types containing authors, papers, terms, and publication venues after data preprocessing and extraction.
Freebase is a huge collaborative knowledge graph, which contains 8 genres of entities.
Amazon is a co-purchase graph. Its nodes represent goods and edges indicate that two goods are frequently bought together.


\textbf{Parameters.} We randomly create motifs to test different situations and reported the average running time, space cost and effectiveness metrics. In particular, we first generate a small HIN by conducting a random walk on the data graph following~\cite{bi2016efficient}. 
After that, we randomly choose a vertex in the small HIN as the target vertex and select the small HIN as a motif if it has at least two vertices with the type of target vertex. 
For performance evaluation, we create motifs with varying sizes from 3 to 7 (default size is 5) because the size of the motifs is bounded from 3 to 7 in real applications~\cite{kleindessner2019guarantees,milo2002network,hu2019discovering}. We randomly create 5 motif sets, each of which contains 100 motifs of the same size.
We treat the running time of a query as infinite (\textbf{Inf}) if the query set cannot be finished in 24 hours.



 \begin{table}[t]
    \caption{Dataset Statistics}
    \centering
    \begin{tabular}{c c c c c}
        \hline
        Dataset &   Vertices   &   Edges   &   Vertex types  &  motifs  \\
        \hline

        IMDB  &   11,616   & 34,212   &   3  & 6 \\
        DBLP  &   26,128   & 239,566   &   4 & 7  \\
        Freebase  &   180,098   & 1,057,688   &   8  & 16 \\
        Amazon  &   1,569,960   &   264,339,468 &   107  & 76 \\
        \hline       
    \end{tabular}
    \label{table:dataset}
 \end{table}

 





\textbf{Algorithms.}
 We evaluate four algorithms in our experiments, namely Baseline, FVA, FVA-M, FVA-L. Baseline is the filter-verify solution discussed in Section~\ref{sec:baseline}. FVA is the filter-verify algorithm with the reducing potential target vertices strategy. FVA-M considers the reducing potential target vertices strategy and message-passing based optimization strategy. FVA-L utilize the reducing potential target vertices strategy, message-passing based strategy and the lower-bound to search community, i.e., the Algorithm~\ref{met:lowerbound}.
 
 

\textbf{Environment.}
All the experiments are implemented in Python 3.7 programming language and are conducted on a Linux system that has an Intel Core i5 CPU @ 2GHz and 8GB of memory. For subgraph isomorphism, we use Grand-Iso~\cite{Matelsky_Motifs_2021}, which is a state-of-the-art subgraph isomorphism algorithm.

\subsection{Evaluation of Efficiency} 
\begin{figure}[t]
\centering

\includegraphics[scale=0.65]{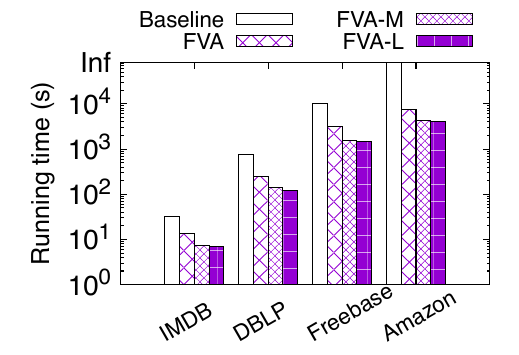}

\caption{Time cost of Target-aware Community Search on Four Datasets under Default Parameter Setting}
\label{fig:default}
\end{figure}

In this subsection, we present the performance of our proposed three algorithms compared with the baseline solution. Figure~\ref{fig:default} demonstrates the time cost of four methods over four datasets under the default parameter settings. Clearly, FVA-L runs much faster than the other three methods for every dataset. In specific, FVA-L reduces the time cost by $\times$2.41, $\times$3.18, and $\times$11.15 compared to the baseline method for IMDB, DBLP, Freebase, and Amazon dataset, respectively.

To show the impact of each parameter, we also evaluate the efficiency of the proposed algorithms over four datasets by varying the motif size $|V_q|$.

\begin{figure*}[htbp] 

\captionsetup[subfloat]{captionskip=-4pt} 

\centering
\hfill

\subfloat[IMDB] {{\includegraphics[scale=0.20]{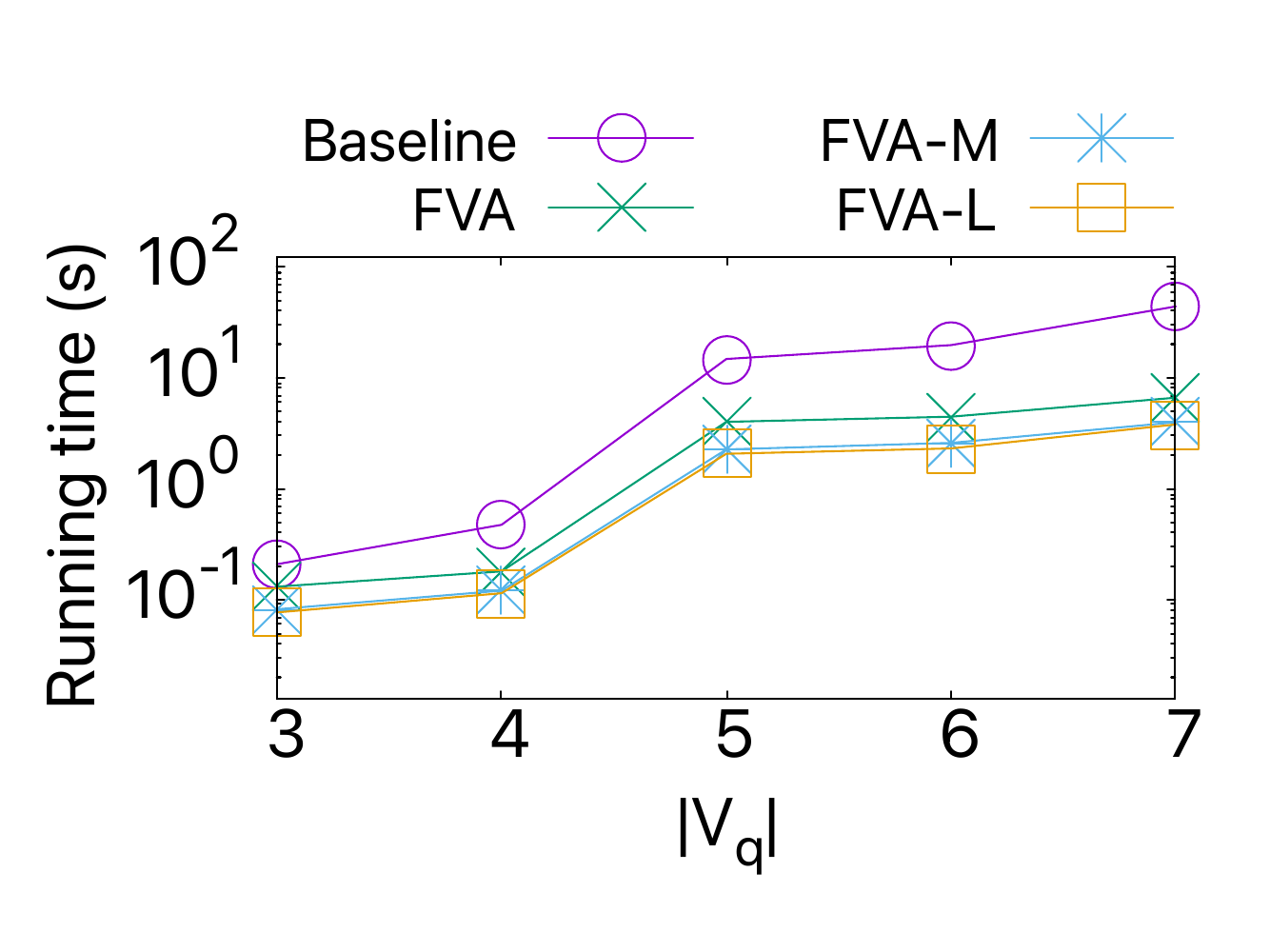}\label{int:IMDB} } }  
\subfloat[DBLP]{{\includegraphics[scale=0.20]{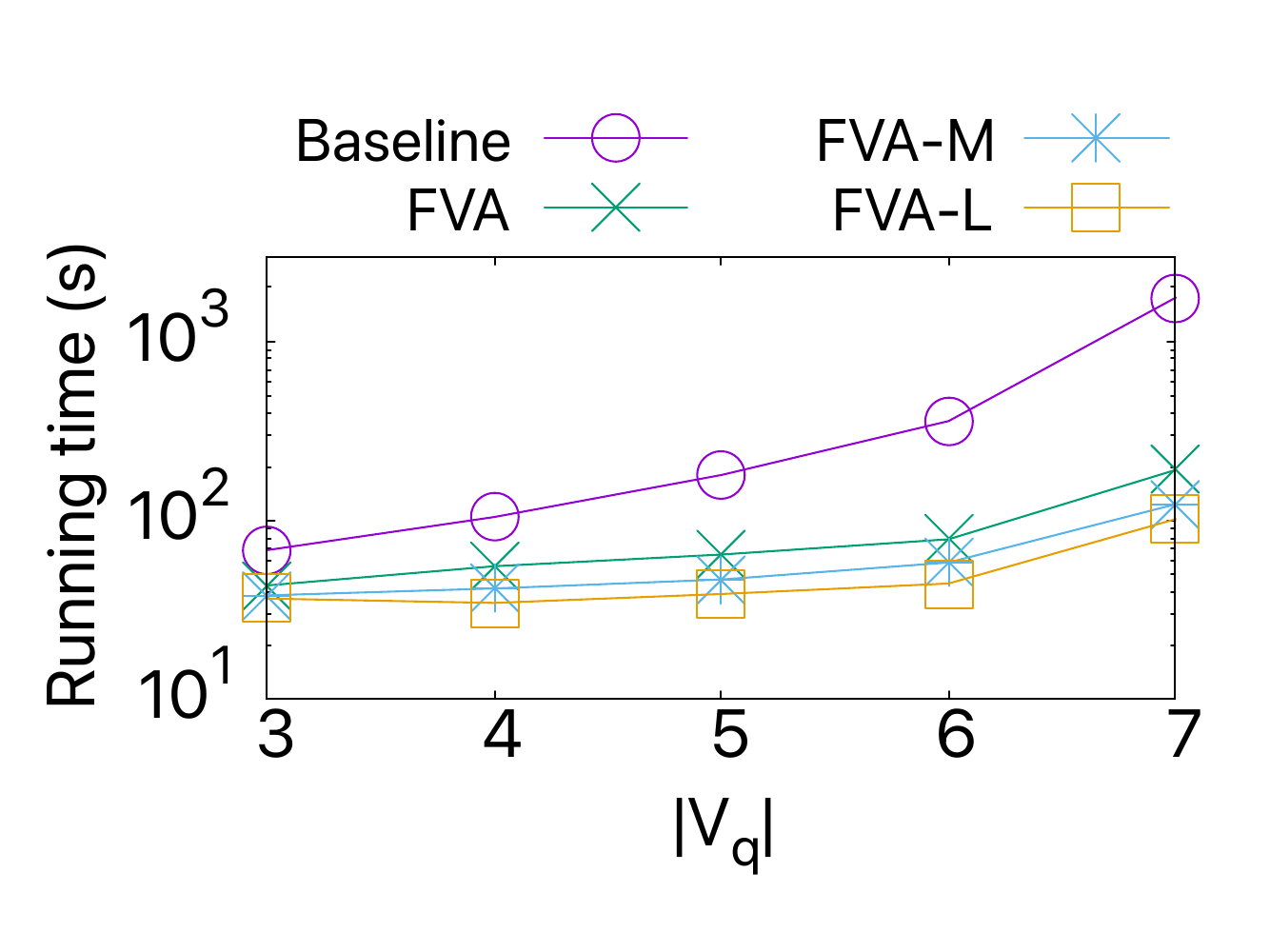}\label{int:DBLP} }} 
\subfloat[Freebase]{{\includegraphics[scale=0.20]{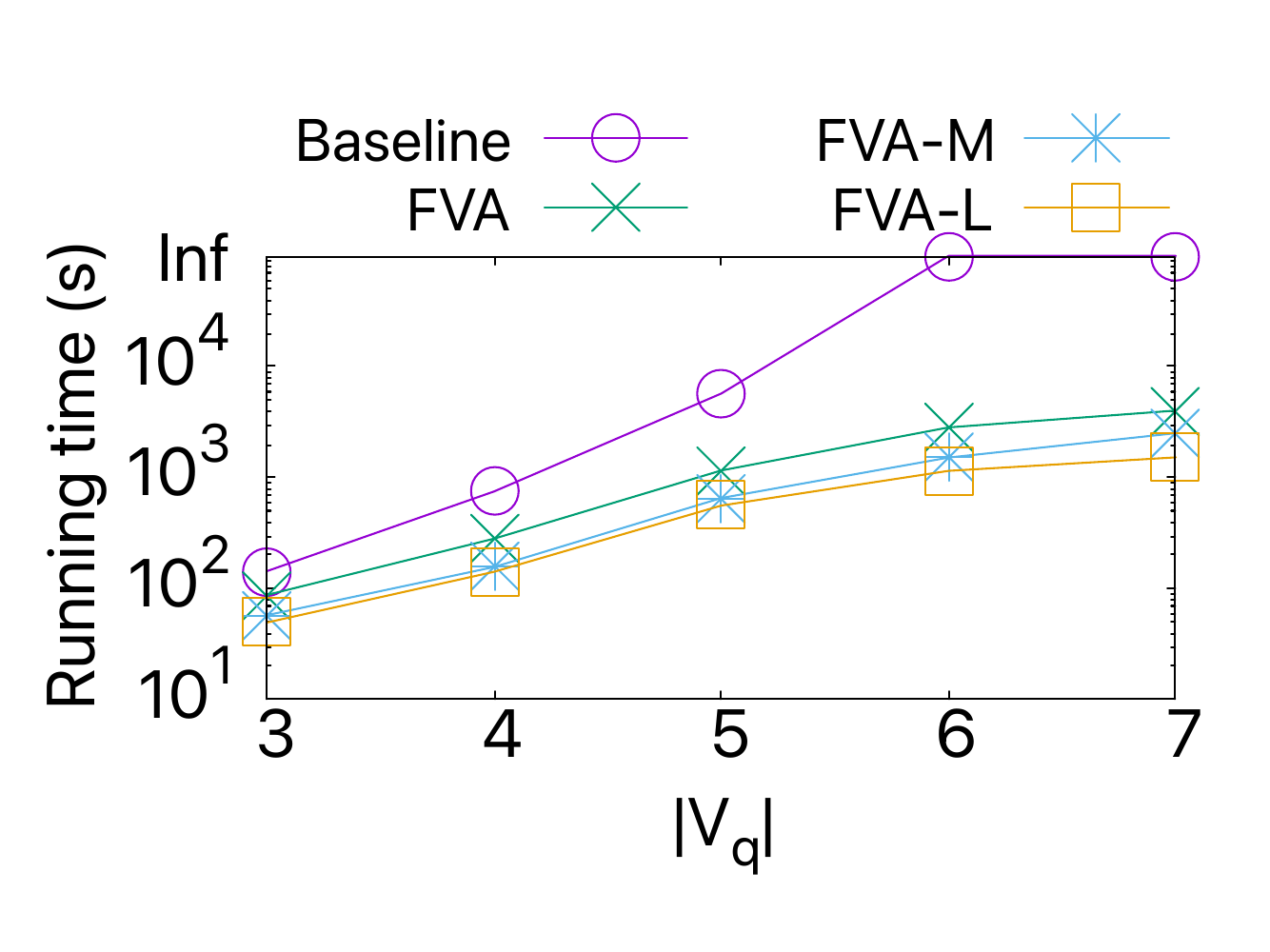}\label{int:Freebase} }} 
\subfloat[Amazon]{{\includegraphics[scale=0.20]{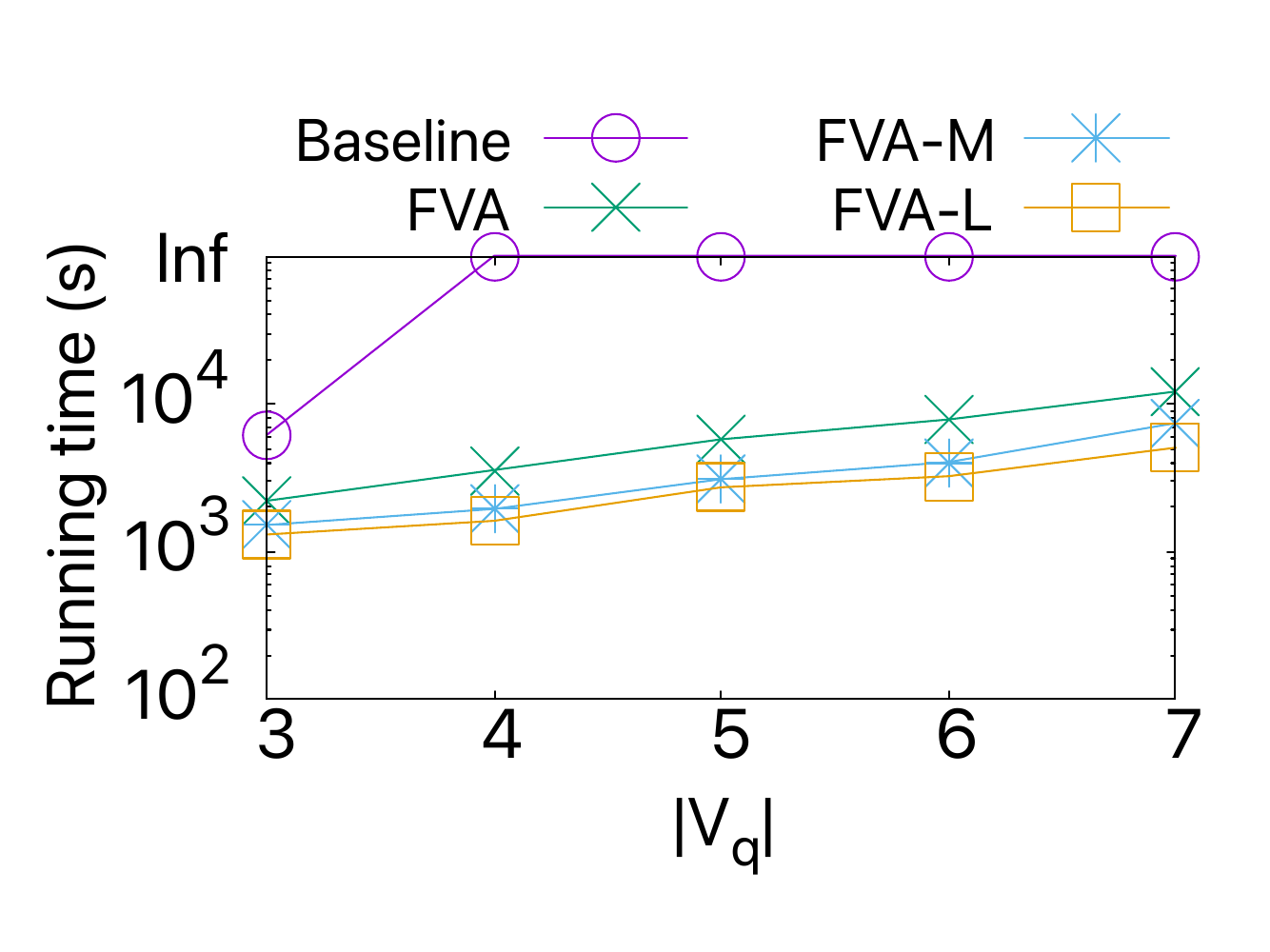}\label{int:Amazon} }} 

\vspace*{-3mm}
\caption{Efficiency evaluation over motif size $|V_q|$ on four datasets}
\label{fig:efficiencyevaluationvaryingvq}
\end{figure*}




\begin{figure*}[htbp] 
\captionsetup[subfloat]{captionskip=-4pt} 
\vspace{-0.4cm}
\centering
\hfill

\subfloat[IMDB]{{\includegraphics[scale=0.20]{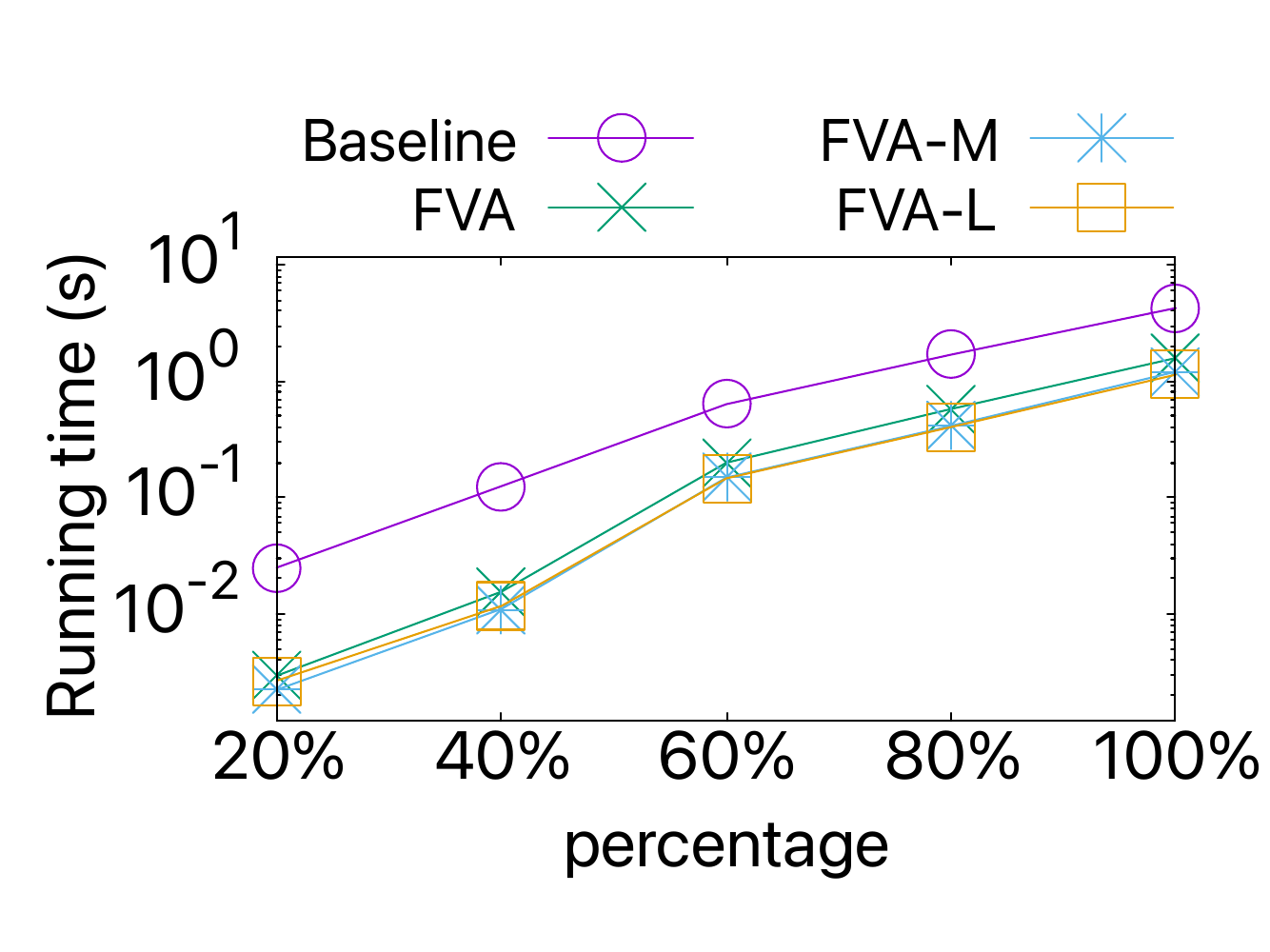}\label{int:IMDB_scability} }} 
\subfloat[DBLP]{{\includegraphics[scale=0.20]{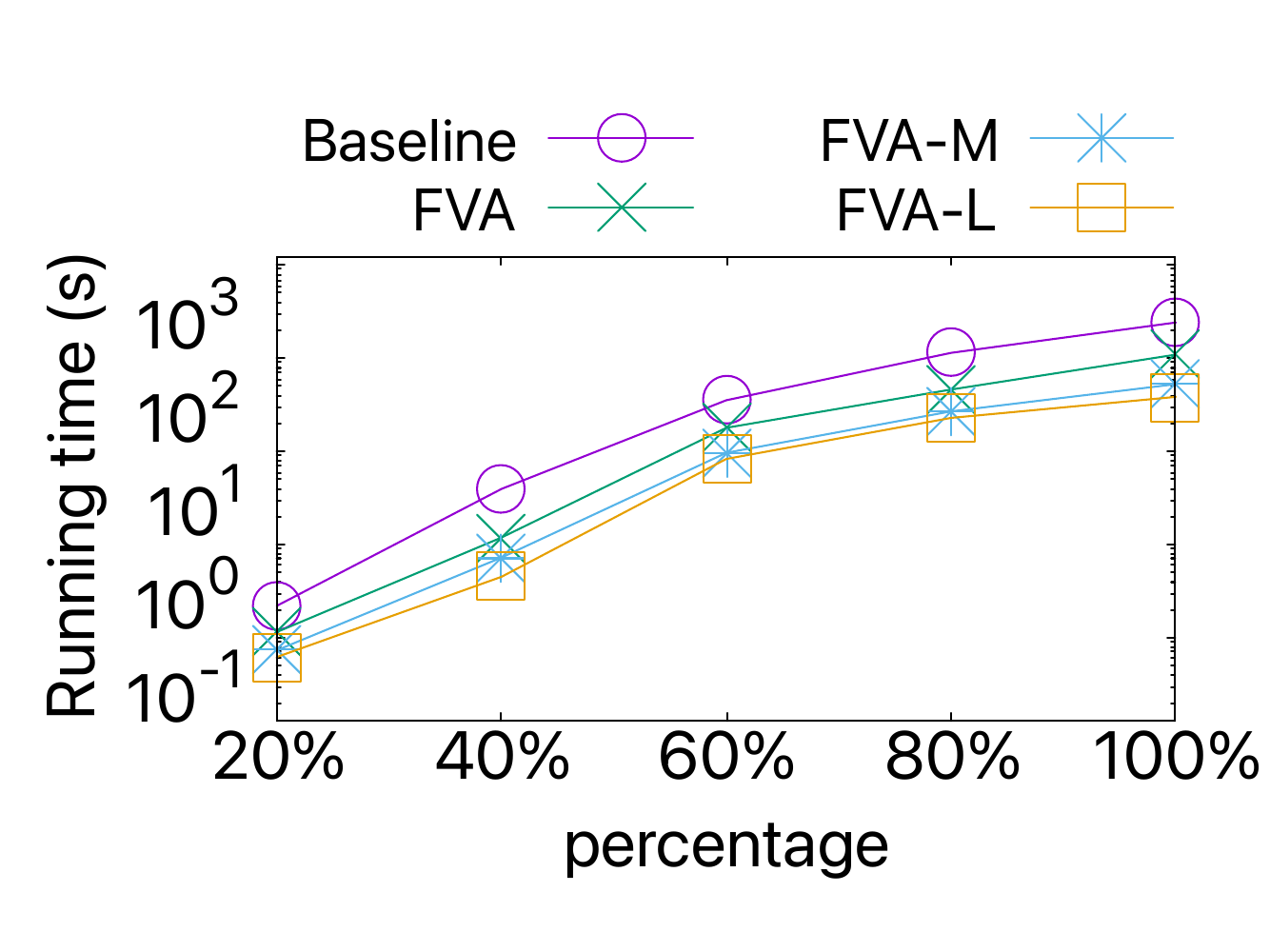}\label{int:DBLP_scability} }} 
\subfloat[Freebase]{{\includegraphics[scale=0.20]{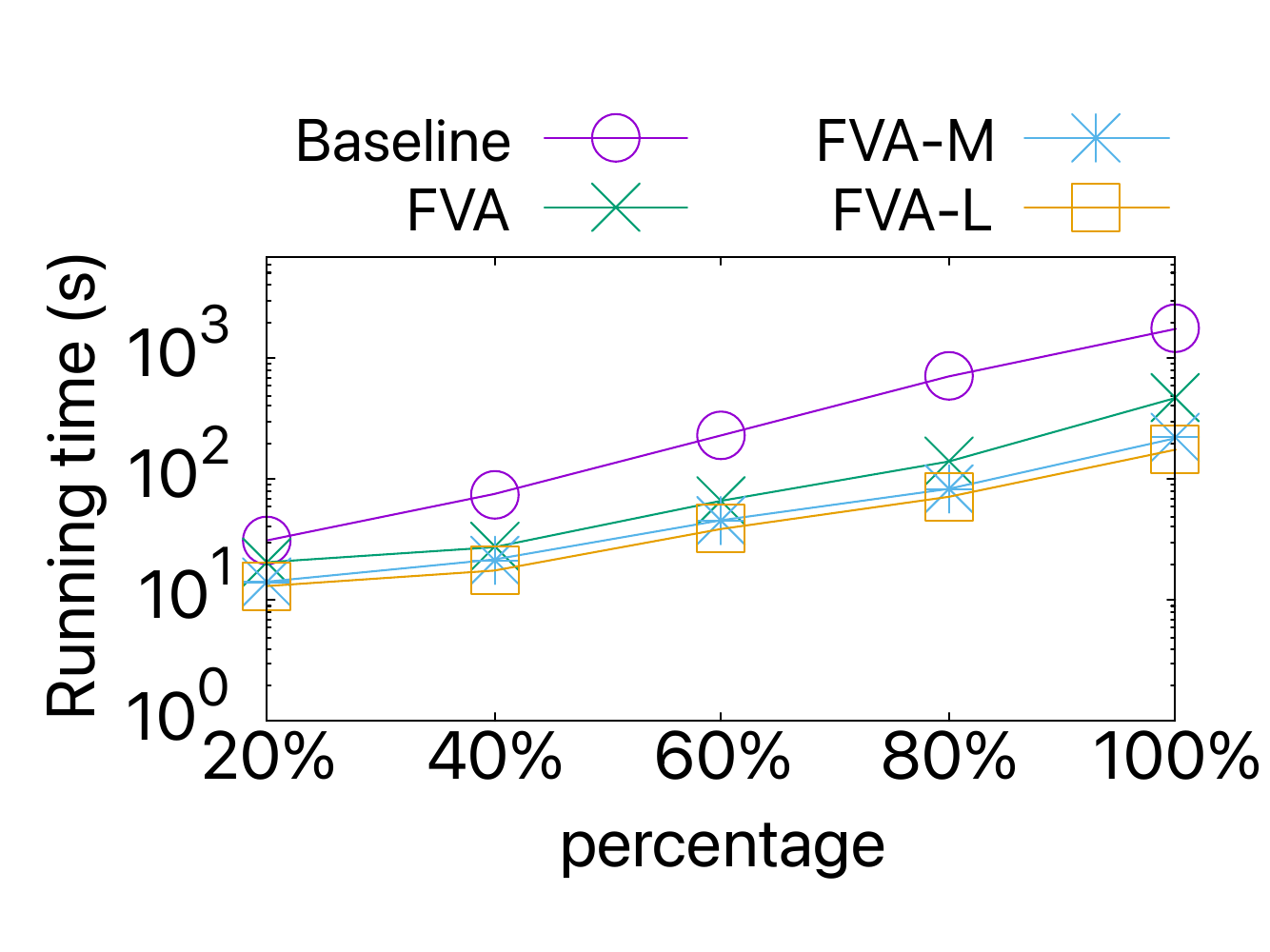}\label{int:Freebase_scability} }} 
\subfloat[Amazon]{{\includegraphics[scale=0.20]{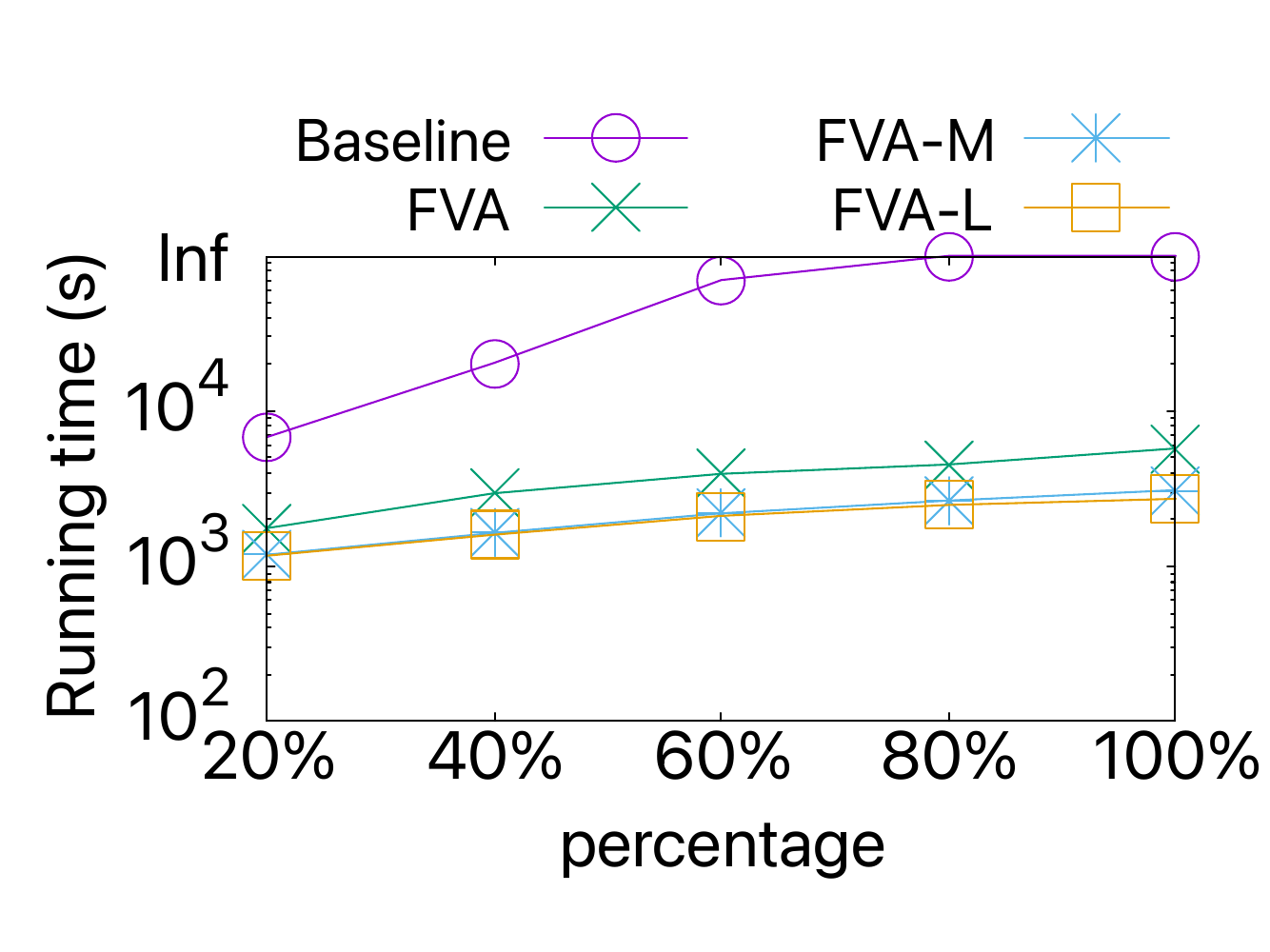}\label{int:Amazon_scability} }} 

\vspace*{-3mm}
\caption{Time cost of four methods with different sampling ratios on four datasets}
\label{fig:scability}
\end{figure*}

\begin{figure*}[htbp]
\captionsetup[subfloat]{captionskip=-2pt} 

\begin{minipage}{0.60\textwidth}
\centering

\subfloat[PathSim]{{\includegraphics[height=3cm, width=4.5cm]{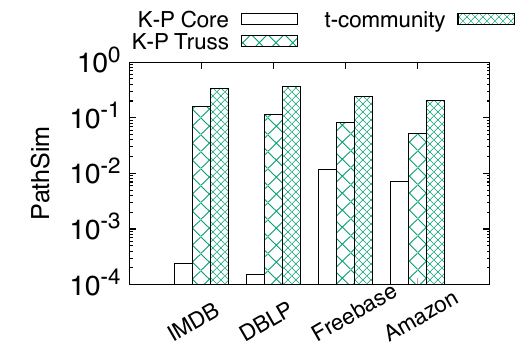}\label{int:SCM} }} 
\subfloat[Density]{{\includegraphics[height=3cm, width=4.5cm]{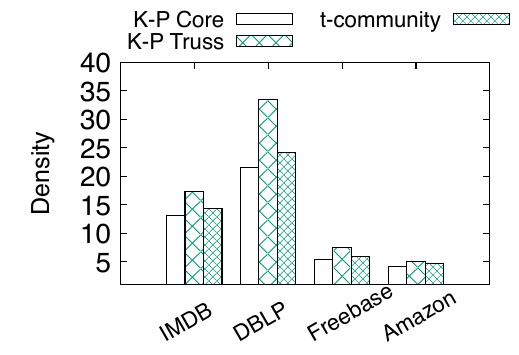}\label{int:Density} }} 
\subfloat[Closeness]{{\includegraphics[height=3cm, width=4.5cm]{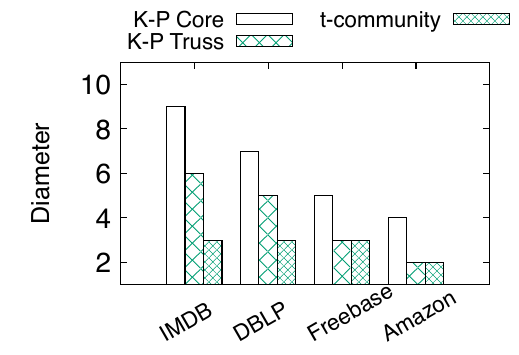}\label{int:Closeness} }}  

\vspace*{-3mm}
\caption{Effectiveness Analysis}
\end{minipage}\hfill
\begin{minipage}{0.23\textwidth}
\centering
\vspace*{2mm}
\includegraphics[height=3.5cm, width=4.5cm]{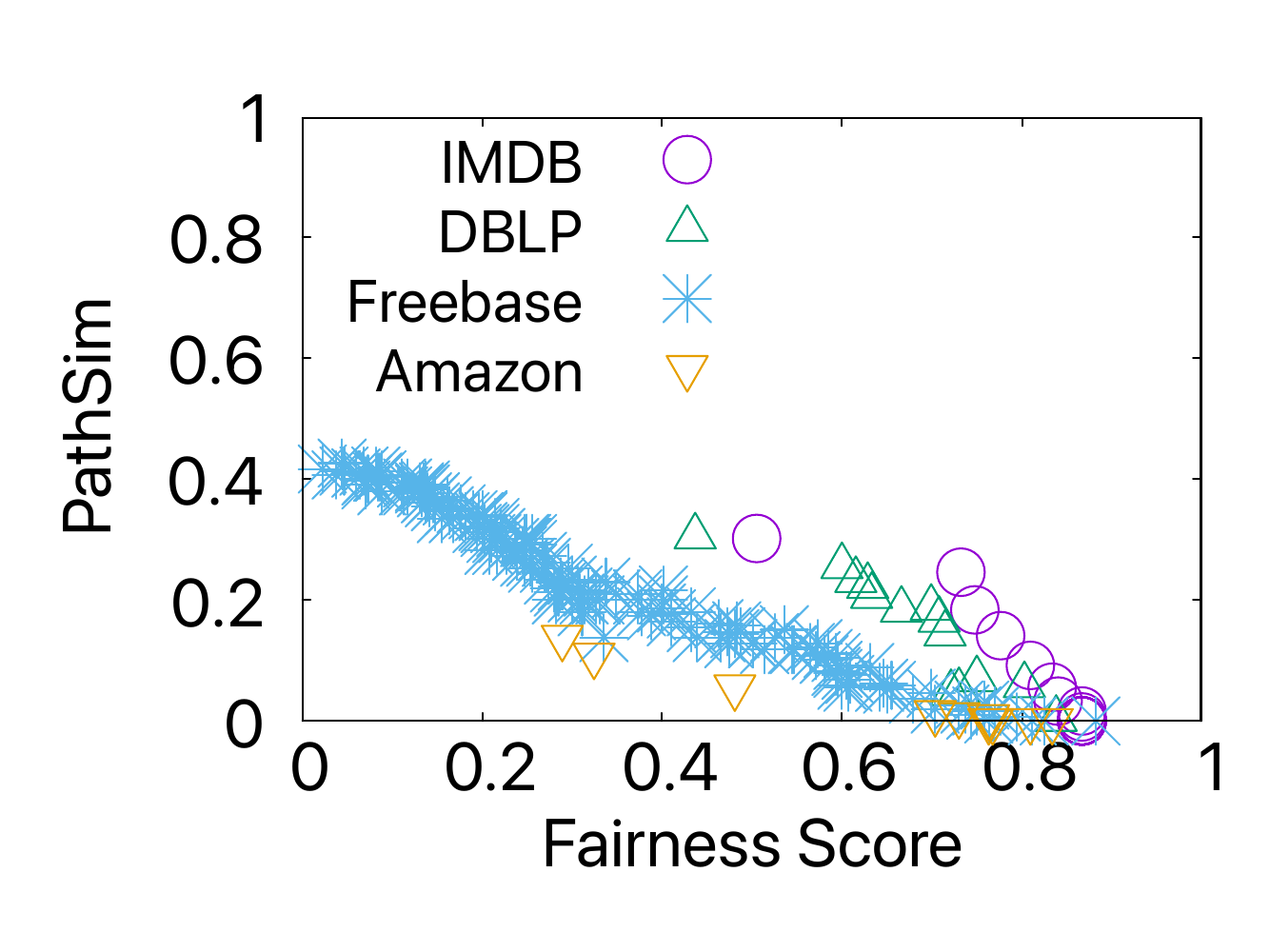}

\vspace*{-3mm}
\caption{Fairness Analysis}
\label{int:evaluation_of_fairness}
\end{minipage}

\end{figure*}

\begin{figure*}[htbp] 
\captionsetup[subfloat]{captionskip=-2pt} 
\vspace{-0.4cm}
\centering
\hfill

\subfloat[IMDB]{{\includegraphics[height=3.2cm, width=4.45cm]{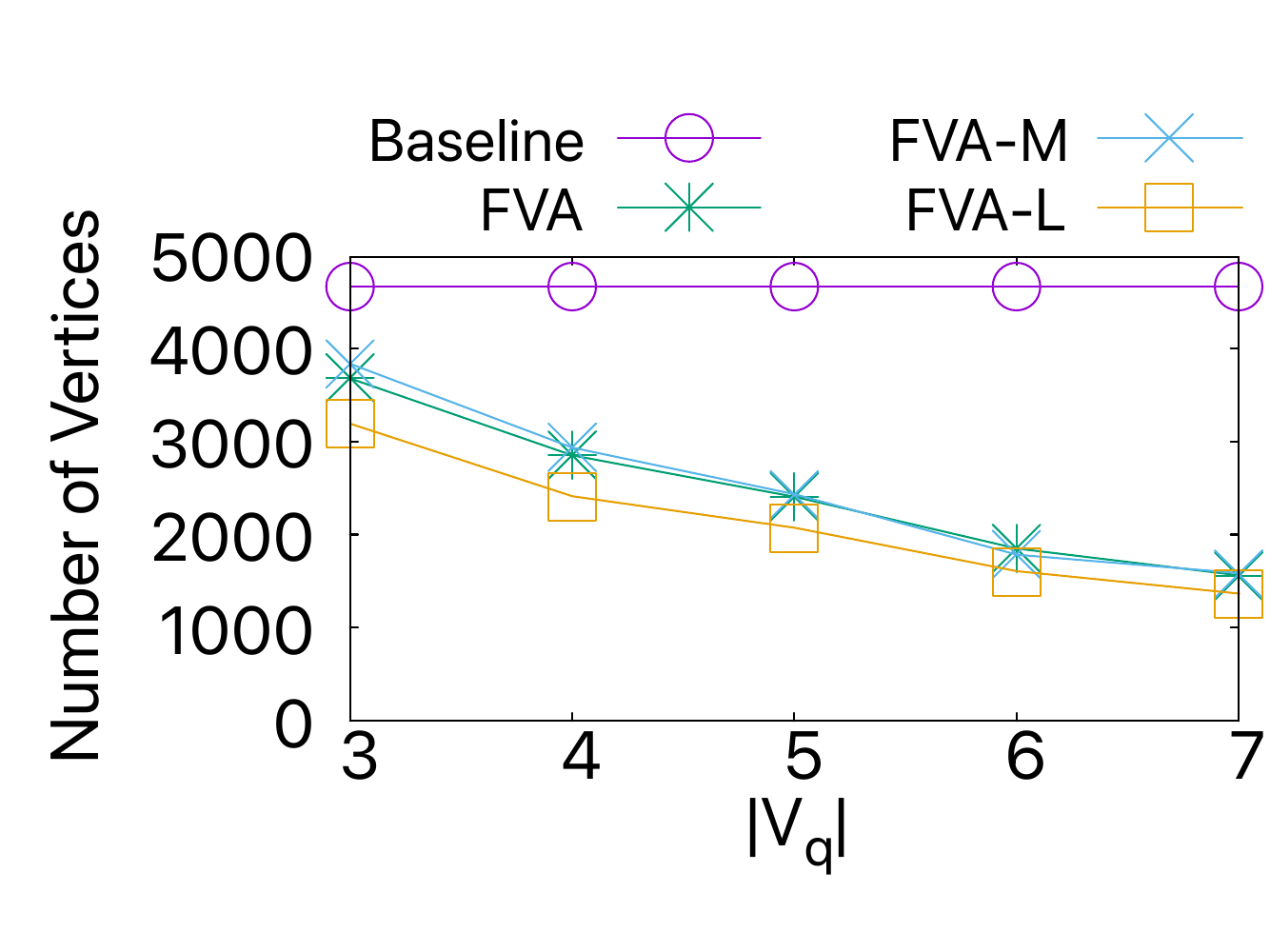}\label{int:DBLP_scability} }} 
\subfloat[DBLP]{{\includegraphics[height=3.2cm, width=4.45cm]{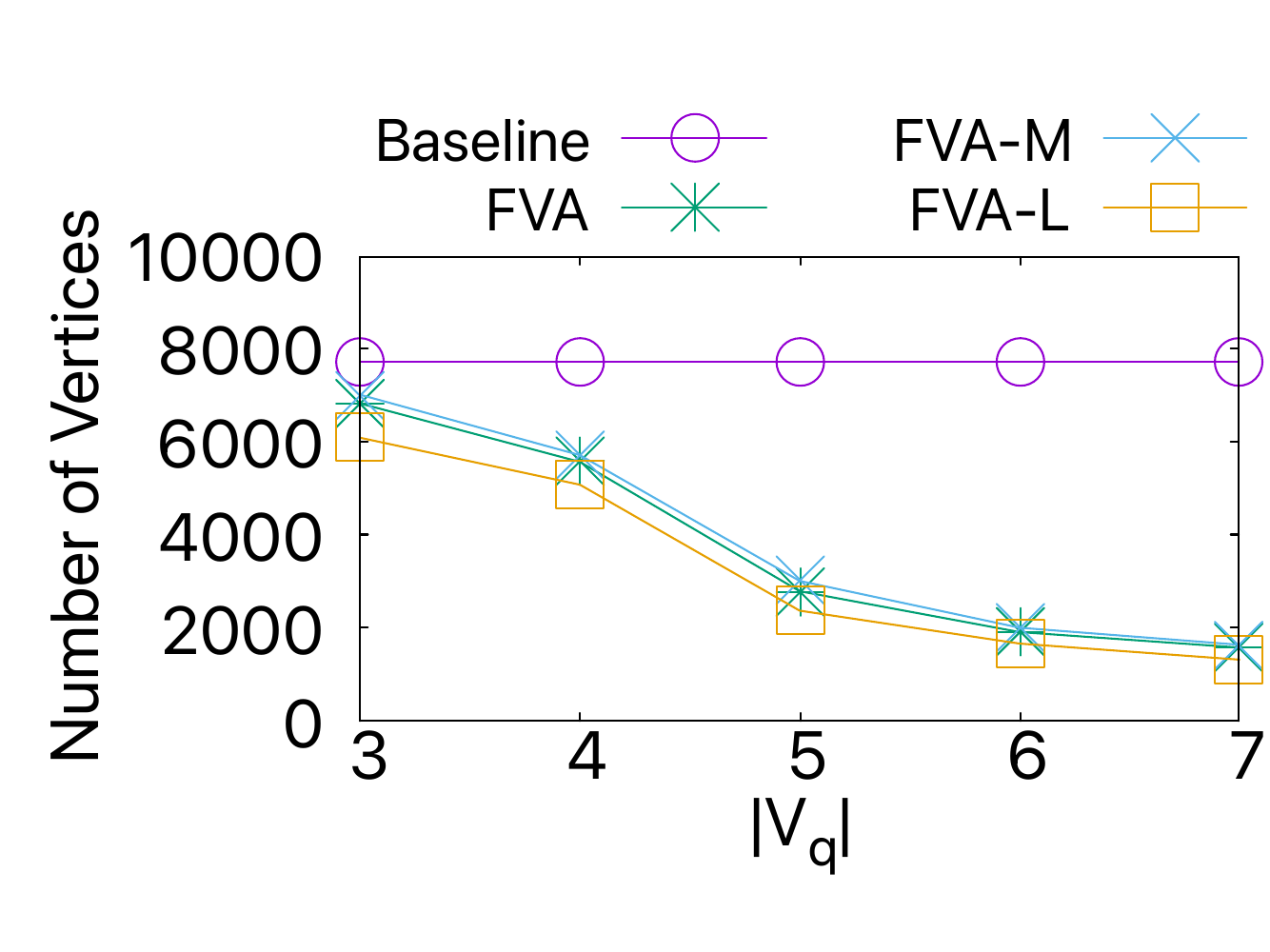}\label{int:DBLP_scability} }} 
\subfloat[Freebase]{{\includegraphics[height=3.2cm, width=4.45cm]{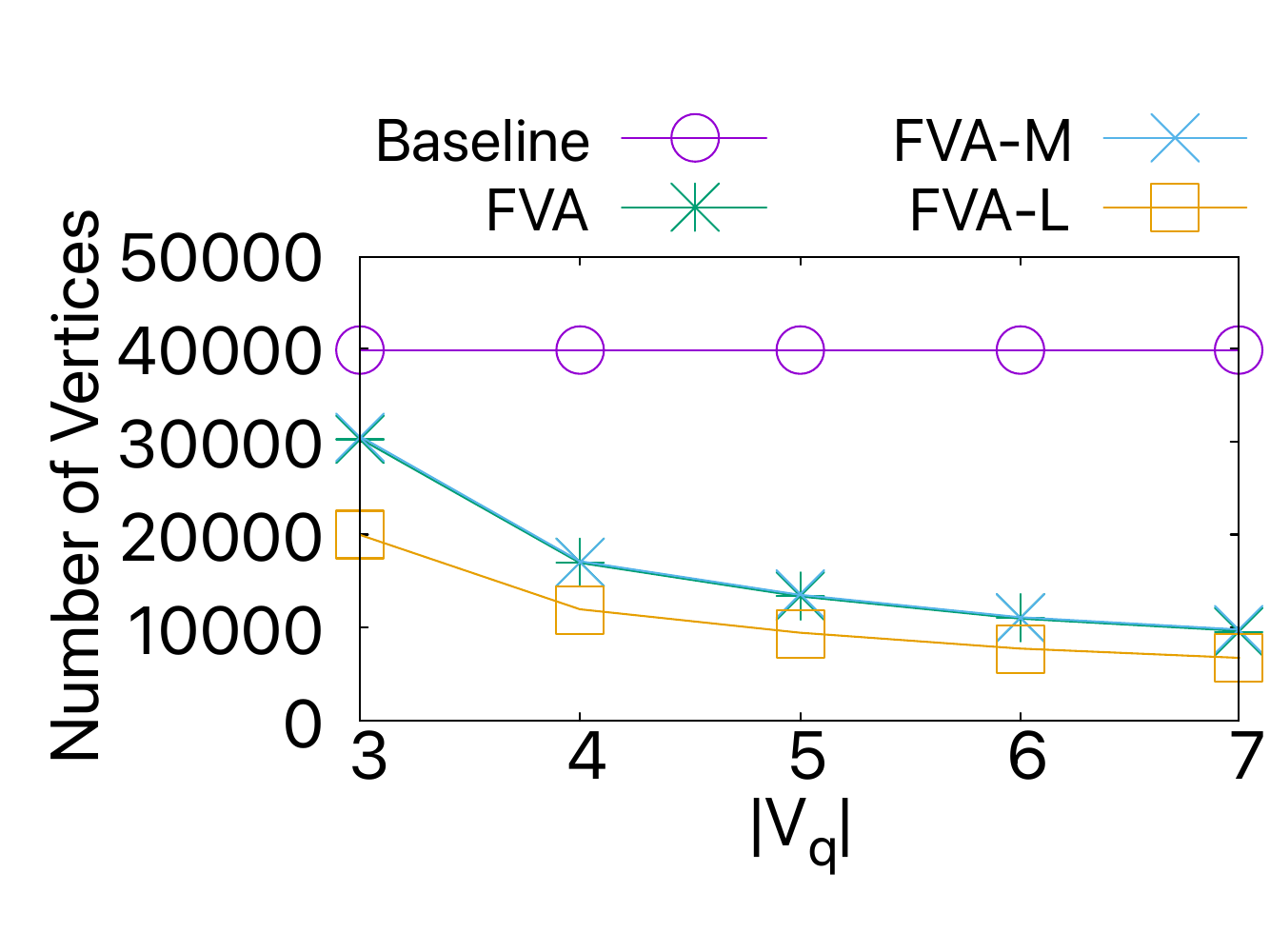}\label{int:DBLP_scability} }} 
\subfloat[Amazon]{{\includegraphics[height=3.2cm, width=4.45cm]{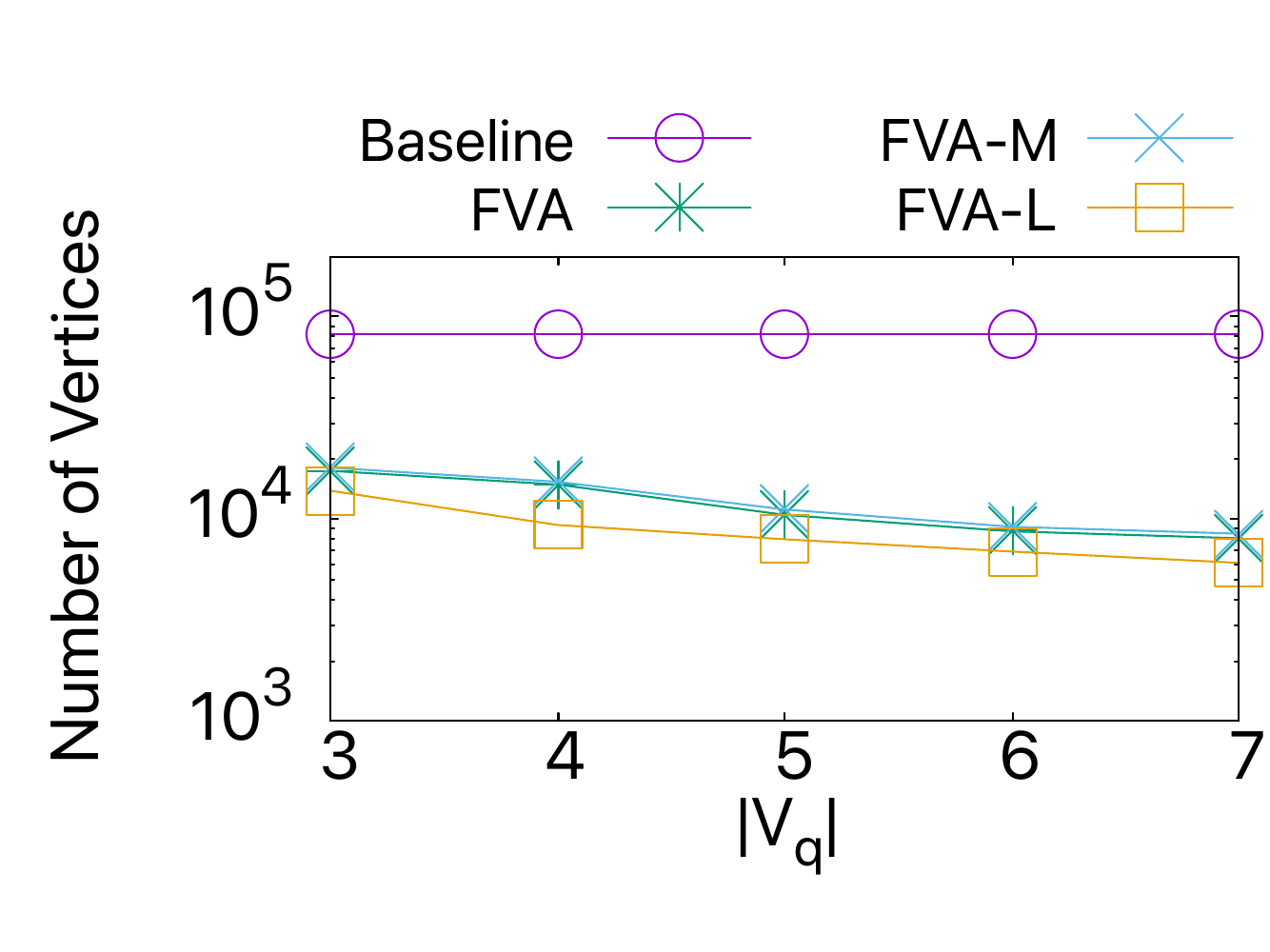}\label{int:DBLP_scability} }} 

\vspace*{-2mm}
\caption{The number of visited vertices after performing baseling, FVA and FVA-L over motif size $|V_q|$}
\label{fig:pruning}
\end{figure*}

\textbf{Varying motif size $|V_q|$.}
Figure~\ref{fig:efficiencyevaluationvaryingvq} shows the average time cost of the four algorithms when motif size $|V_q|$ varies from 3 to 7. We obverse all the algorithms consume higher time costs when the motif size increases. This is because the cost of active level calculation (i.e., motif enumeration) increases with the motif size. In addition, the three proposed algorithms consume significantly less time than the baseline method and consume decreasing time in a similar trend as the results under the default parameter settings. Thus, we conclude that our proposed methods effectively enhance the performance of target-aware community search under different motif sizes.


\textbf{Memory overheads analysis.}
Table~\ref{tab:memory} shows the total memory overheads of the methods on four datasets. Note that the memory overheads of algorithms do not contain the memory overhead of the subgraph isomorphism algorithm. Obviously, the memory overheads of FVA, FVA-M, and FVA-L are significantly higher than that of baseline. This is because FVA, FVA-M, and FVA-L store the candidate vertices of each query vertex in the exploration-based filter stage. Additionally, we also see the memory overheads of FVA, FVA-M, and FVA-L are almost the same. This is because the message-passing based postponing enumeration search and lower bound-based approach changed the method of searching, which does not incur additional memory overhead. Furthermore, we can see the memory overheads of same method in Amazon dataset is significant lower than other three datasets. This is because the number of motif instances and target-aware community stored in memory is less than other three datasets.

\begin{table}[t]
    \caption{Evaluation of memory overheads (MB)}
    \vspace{5pt}
    \centering
    \begin{tabular}{c | c | c | c | c}
        \hline
        Dataset &   Baseline   &   FVA & FVA-M & FVA-L    \\
        \hline

        IMDB & 4.15  &   11.10   &  11.11 & 11.13  \\ \hline

        DBLP & 8.22  &   34.41   &  34.44 & 34.43  \\ \hline

        Freebase & 7.16  &   27.38   &  27.41 & 27.47  \\ \hline

        Amazon & 4.71  &   9.31   &  9.32 & 9.32  \\ \hline
 
        \hline

        \hline        
    \end{tabular}
    \label{tab:memory}
\end{table}

\begin{figure}[t]
\centering
\vspace*{-4mm}

\subfloat[IMDB]{{\includegraphics[scale=0.19]{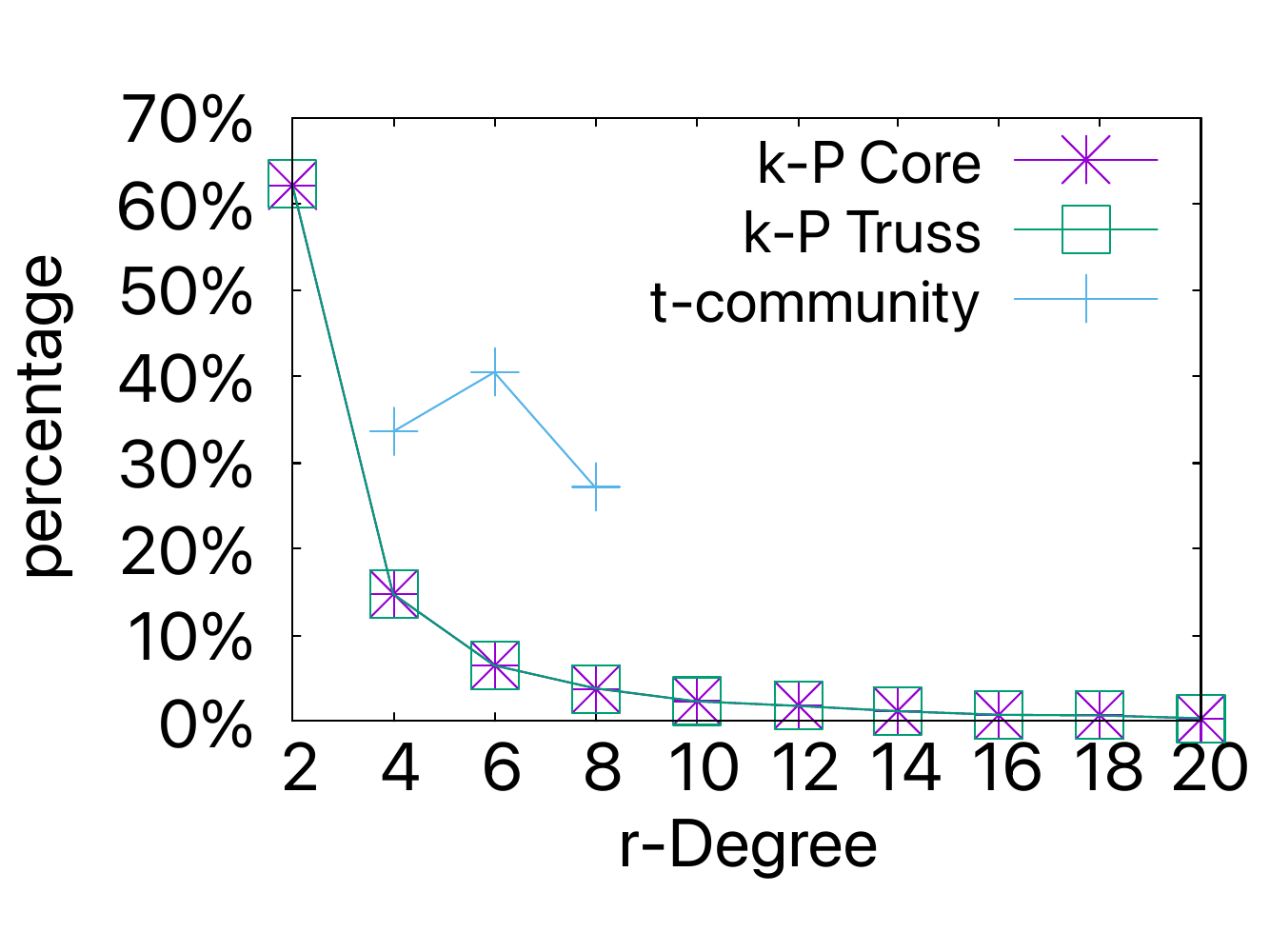}\label{int:imdb_degree} }} 
\subfloat[DBLP]{{\includegraphics[scale=0.19]{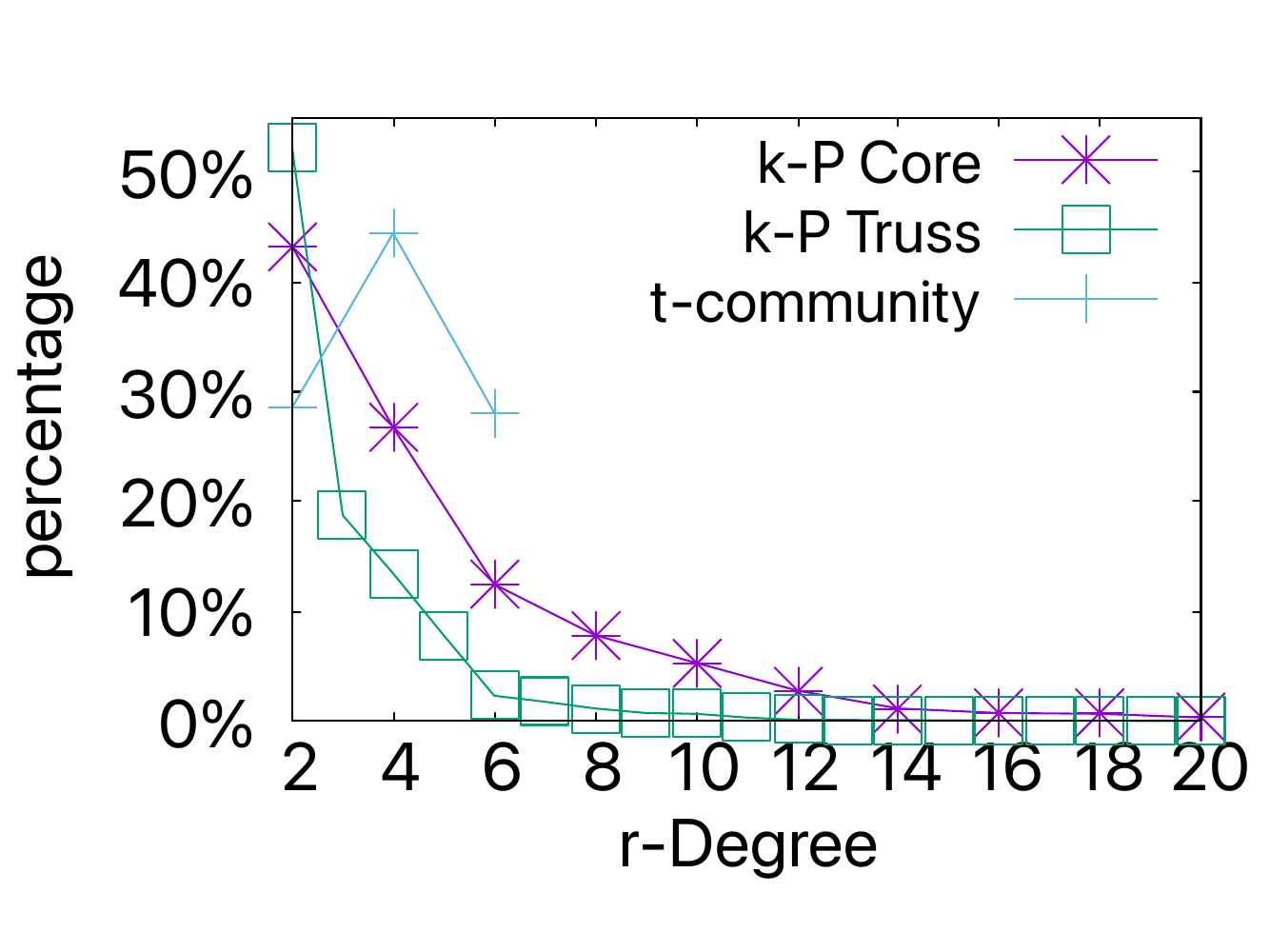}\label{int:dblp_degree} }} 

\caption{r-degree distribution of communities}
\label{fig:r-degree}

\end{figure}

\textbf{Efficiency of pruning strategies.} Figure~\ref{fig:pruning} evaluates the pruning efficiency of Baseline, FVA, FVA-M and FVA-L by comparing the number of remaining vertices that need to calculate their active levels on four datasets with varying $|V_q|$. As can be seen from Figure~\ref{fig:pruning}, FVA, FVA-M, and FVA-L can significantly reduce the number of vertices compared to the baseline method as expected. Moreover, the number of remaining vertices decreases as $|V_q|$ increases. For instance, in the IMDB dataset, FVA reduces the number of vertices from 4670 to 1592; FVA-M reduces the number of vertices from 3683 to 1587; and FVA-L further reduces the number of vertices from 3198 to 1372. Compared with the baseline method, FVA and FVA-M have nearly same number of remaining vertices. These results confirm that the pruning effect of FVA and FVA-M mainly comes from the exploration-based filter. In addition, we can find that FVA-L substantially reduces the number of vertices compared to FVA. This is because FVA-L not only prunes the unpromising vertices in the exploration-based filter but also prunes the vertices of the candidate target-aware communities that do not pass the lower bound of fairness score.

\subsection{Evaluation of Scalability}
We evaluate the scalability of four proposed algorithms over datasets IMDB, DBLP, Freebase and Amazon. For each dataset, we generate four small datasets with different sizes by randomly sampling 20\%, 40\%, 60\%, and 80\% vertices from the datasets, respectively. Note that the dataset itself is considered with the 100\% data size. Figure~\ref{fig:scability} shows the time cost of four algorithms on the size-varying datasets. With the increase of the dataset size, we can observe that the running time of Baseline, FVA, FVA-M and FVA-L has a linear increasing trend. This implies that our proposed three algorithms are easily applied to large-scale networks. However, the baseline method can not be finished within 24 hours in large datasets (i.e., Amazon and Freebase). Thus, we conclude that the baseline solution has limitations in scalability.

\subsection{Evaluation of Effectiveness}
To show the effectiveness of community search in HIN, we compare our fairest target-aware community (t-community) with the $k$-$\mathcal{P}$ Core~\cite{fang2020effective} and $k$-$\mathcal{P}$ Truss~\cite{9101354}. To achieve this, we 
generate five symmetric meta-paths $\mathcal{P}$ discussed in~\cite{fang2020effective} that are the sequences of vertex types between two given target vertex types, and can be seen as the motifs in our work. We calculate the quality metrics of each $k$-$\mathcal{P}$ Core and $k$-$\mathcal{P}$ Truss, and report the highest metrics as result. We utilize the following metrics to analyze the quality of communities.

\textbf{Relational Degree of Community Members.} Conventionally, the degree of a vertex is the number of edges connecting it. To adapt it for communities in HINs, we redefine it as the number of $M$-neighbors of a vertex and call it relational degree (r-degree). For each community, we count the percentages of vertices whose r-degree varies from 1 to 20. Due to the space limitation, we only report the average percentage values on IMDB and DBLP datasets in Figures~\ref{fig:r-degree}. Clearly, compared to $k$-$\mathcal{P}$ Core and $k$-$\mathcal{P}$ Truss, the variances of the r-degree of target-aware community are smaller. Thus, the engagement between vertices in t-communities is more similar than the engagement between vertices in $k$-$\mathcal{P}$ Core and $k$-$\mathcal{P}$ Truss.

\textbf{Similarity of Community Members.} We measure the similarity of community members by using PathSim~\cite{wang2020howsim}. Specifically, we first find communities of $k$-$P$ core, $k$-$P$ Truss and target-aware community, then compute the PathSim value for each pair of vertices in these communities. Figure~\ref{int:SCM} shows the average PathSim values on four datasets. Clearly, target-aware communities achieve higher similarity values than those of $k$-$P$ core and $k$-$P$ Truss, so their members are more similar to each other.

\textbf{Density of link relationships.} To measure the density of link relationships, we extend the traditional density and redefine it as the number of vertex pairs that are connected by meta-path over the number of vertices in community. The average densities for communities of each community model are depicted in Figure~\ref{int:Density}. We observe that the densities of t-communities are higher than the $k$-$P$ core but lower than the $k$-$P$ Truss. Thus, the target-aware community model achieves stronger cohesiveness than the $k$-$P$ core model but less cohesiveness than the $k$-$P$ Truss model.

\textbf{Closeness of Community.} To measure the closeness of communities, a commonly-used metric is the diameter~\cite{7930032}, which is the largest shortest distance between any pair of vertices in the community. To adapt it for communities in HINs, we redefine the distance as motif-constrained distance, i.e., $M$-distance; that is, the $M$-distance between two target vertex instances linked by an instance of the motif is 1. In our experiment, we first calculate the shortest $M$-distance of each pair of target vertex instances in these communities and report the largest distance in Figure~\ref{int:Closeness}. Clearly, the t-communities have smaller diameter than $k$-$P$ cores and $k$-$P$ Trusses on all datasets, which means that the community members tend to have closer relationships.

\textbf{Evaluation of Fairness.}
We conduct the study on the relationship between fairness scores and the similarity between members in t-communities. Specifically, we first find t-communities on each dataset, then compute the fairness score of each t-communiy and the PathSim value for each pair of vertices in each community. Figure~\ref{int:evaluation_of_fairness} shows the average PathSim values against the fairness scores of t-communities on four datasets. Clearly, the average PathSim consumes higher when the fairness score decreases. For instance, we can find 14 t-communities in DBLP, the fairness scores are 0.16, 0.19, 0.25, 0.26, 0.27, 0.28, 0.29, 0.3, 0.33, 0.36, 0.37, 0.38, 0.4, 0.56, and the average PathSim values are 0.36, 0.25, 0.23, 0.22, 0.20, 0.18, 0.18, 0.16, 0.14, 0.06, 0.05, 0.04, 0.04, 0.005, respectively. Thus, we conclude that the community members will be more similar if the target-aware community has a higher fairness score.




\section{Related Work} \label{sec:relatedwork}
\textbf{Community search:} Community search aims to query cohesive subgraphs that satisfy the customized query request. To measure cohesiveness of a subgraph, existing works develop different community models such as $k$-core~\cite{batagelj2003m, wood2015minimal}, k-truss~\cite{cohen2008trusses}, k-clique~\cite{kumpula2008sequential}, k-edge-connected component~\cite{chang2013efficiently}, and the k-plex~\cite{conte2017fast}. However, these works focus on searching communities over homogeneous graphs, which cannot be directly used in a heterogeneous network because the relation between vertex types is different. Recently, researchers attempted to find cohesive communities from HINs. For instance,~\cite{fang2020effective, 9101354, DBLP:journals/pvldb/JiangFMCL22,DBLP:journals/pvldb/JianWC20,hu2019discovering} utilize different customized query requests such as meta-path~\cite{fang2020effective, 9101354, DBLP:journals/pvldb/JiangFMCL22}, relational constraint~\cite{DBLP:journals/pvldb/JianWC20} and motif~\cite{hu2019discovering} to extend traditional community models. However, these studies did not consider the notion of fairness, which may lead systematic discrimination for disadvantaged people in communities. 

\noindent\textbf{Fairness-aware Community Mining:} The notion of fairness graph retrieval has received much attention in recent years. It is used to alleviate the bias problem caused by the tendency of retrieval. 
For example,~\cite{kleindessner2019guarantees} proposed a fair spectral clustering algorithm to generate communities, which ensures that each cluster contains roughly the same number of group elements. 
~\cite{kamishima2012enhancement, yao2017beyond} provided a parity-based fairness measurement to distinguish the differences in behavior between dominant and disadvantaged users. 
~\cite{fu2020fairness} proposed a heuristic reordering based fairness algorithm to reduce the influence of active users' history on inactive users' recommendations. 
~\cite{DBLP:conf/iclr/LiWZHL21} refined the attributes that need to be considered fairly. 
It applied the fairness measures to user-defined attributes and allowed other attributes to bias in the recommendation.
However, the current works focus on how to make unbiased recommendations through keeping the similarity of certain metrics between groups, which can not make sure the fairness in fine granularity level (e.g., keep fairness between members in a group).


\section{Conclusions} \label{sec:conclusion}

In this paper, we first discussed the necessity of individual fairness in community search problem, then formalized the novel problem of individual fairest community search, which considers the similarity of active level between members in a community. To model the relationships between members of communities and customized query requests of users, we extended the well-known concept of motif. To tackle this problem, we first proposed an filter-verify algorithm, then we propose an exploration-based filter strategy to reduce the potential target vertices. Based on the filter, we developed a message-passing based postponing enumeration search method to reduce redundant computation. We further boosted the query efficiency by identifying and pruning the unfair community in advance during the process of community search. Our experimental results on four real HINs show the efficiency of our proposed filter and algorithms, and the effectiveness of our proposed community.


\bibliographystyle{ACM-Reference-Format}
\bibliography{main}
\end{document}